\documentclass[preprint,3p,twocolumn]{elsarticle}
\usepackage{lineno}
\usepackage{amsmath}

\journal{Nuclear Instruments and Methods Section A}

\begin{document}

\begin{frontmatter}

	\title{Online energy discrimination at DAQ front-end level on pixelated TOF-PET systems}

	\author[bit,cib]{Carlos Zorraquino\corref{cor1}}
	\ead{carlos.zorraquino@upm.es}

	\author[ps]{Luis Ferramacho}

	\author[ps]{Ricardo Bugalho}

	\author[lu]{Milan Zvolsky}

	\author[lip]{Tahereh Niknejad}

	\author[lip,ps]{Jose C. Silva}

	\author[ps,vu]{Stefaan Tavernier}

	\author[bit,cib]{Pedro Guerra}

	\author[lip,ps]{Joao Varela}

	\author[bit,cib]{Andres Santos}

	\cortext[cor1]{Corresponding author. Tel.: +34 915495700 ext. 4234. Address: Esc. Ingen. Telecomunic. Avd.Complutense nº 30- Desp.C-203.1,28040- MADRID}

	\address[bit]{Biomedical Image Technologies Lab, Universidad Polit\'ecnica de Madrid, Av. Complutense 30, 28040 Madrid, Spain}
	\address[cib]{CIBER-BBN - Centro de Investigaci\'on Biom\'edica en Red en Bioingenier\'ia, Biomateriales y Nanomedicina, Av. Monforte de Lemos, 3-5.28029 Madrid}
	\address[ps]{PETsys Electronics, Taguspark, Edificio Tecnologia I, 26, 2740-122 Oeiras, Portugal}
	\address[lu]{Institute of Medical Engineering, University of L\"ubeck, L\"ubeck, Germany}
	\address[lip]{LIP - Laborat\'orio de Instrumenta\c{c}\~ao e F\'isica Experimental de Part\'iculas,
		Av. Elias Garcia 14-1, 1000-149 Lisboa, Portugal}
	\address[vu]{Vrije Universiteit Brussel, Brussel, Belgium}

	\begin{abstract}
		Pixelated positron emission tomography (PET) systems produce higher count rates than traditional block detectors because they integrate several detecting channels per detector module. An increased data flow from the detectors leads to higher bandwidth requirements. We aim to optimize the bandwidth usage efficiency by on-the-fly filtering of detected events for non-valid energies. PET systems with a SiPM-ASIC readout scheme are being extensively used to obtain enhanced images on time-of-flight PET scanners. Such digital readout systems are especially interesting for the application of online processing techniques given the ease of access to the  digital information of each detected event. This study analyzes online processing techniques at the DAQ front-end level (on-detector electronics) for pixelated PET systems with SiPM-ASIC readout. In particular, we worked with a tunable online energy discriminating stage. For the optimization of its hardwired internal limits, we analyzed the system energy space. We explored various solutions, some of which are dependent on the system's energy calibration, whereas others were not. Results obtained through the application of various filter versions confirm the minimal resources consumption of such processing techniques implemented at the DAQ front-end level. Our experiments demonstrate how the filtering process reduces the bandwidth needs by excluding all non-valid energy events from the data stream, thus improving the system sensitivity under saturation conditions. Additionally, these experiments highlight how setting proper energy limits can ensure preservation of the system performance, which maintains its original energy and time resolution. In light of these findings, we see great potential for the application of online processing techniques for time-of-flight PET at the DAQ font-end level (on-detector electronics); therefore, we envisage the development of more complex processing methods.

	\end{abstract}


\end{frontmatter}


\section{Introduction}\label{sec:intro}

The state-of-the-art in positron emission tomography (PET) scanner shows a tendency towards pixelated detectors rather than monolithic ones. Pixelated systems have the potential to achieve better spatial resolution than traditional monolithic detectors because the physical size of the detector element usually plays a dominant role
in determining resolution\cite{MOSES2011S236}. However, the use of highly pixelated scanners integrating thousands of detecting channels implies grater system complexity. The higher the degree of pixelation in the detector the more readout channels are required, which increases the cost and complexity of data acquisition (DAQ) \cite{7843591}. In pixelated systems, the DAQ must handle higher detector output data rates, and an uncontrolled aggregated data flow from thousands of detecting channels could saturate the system. Traditionally, the workstation collecting all the events produced by the system performs an offline event energy discrimination. This solution reduces the  data processing computational costs by cutting down the volume of data to be analyzed. Nevertheless, applying this technique on a highly pixelated and high-count rate system does not ease the congestion issue. In these cases, the volume
of data transferred from the detector electronics to the acquisition workstation
tends to saturate the link between detector and data collector. Applying no
particular policy to discriminate and discard events, under saturation, the system dumps any event regardless of the information it contains, which results in
reduced system sensitivity. New online techniques are appearing in the
literature to address this situation. Online energy discrimination, applied
for event rejection when events' energy lays outside the valid PET energy
window, is one of these techniques. Online discriminating techniques can reduce the volume of data transferred without a negative effect on system performance \cite{Wu2013, 8069574, 5076010, 7100955}.

Given its system characteristics, the EndoTOFPET-US scanner \cite{Aubry2013, 10.1007/978-3-319-00846-2_112} can take advantage of the application of these online discriminating procedures. The EndoTOFPET-US is an asymmetric PET scanner with a main detector containing 4096 detecting channels composed of a lutetium-yttrium oxyorthosilicate (LYSO) crystal pixel coupled to a silicon photo-multiplier (SiPM) channel. Simulations predict that this 4K-channel detector should produce a count rate of 40 MHz \cite{Frisch2013}, while real acquisitions show that the practical bandwidth (BW) limit on the system goes down to 25 MHz. Given the small size of the opposite detector, which produces no more than 200 kHz, when the system works in coincidence mode, the data rate will be below the saturation point. However, when the system is operated in singles mode for detector calibration, the bigger detector will saturate the system. Experiments with this equipment have shown that the main detector's data stream contains not only valid gamma events but a large proportion of low-energy events (noise events) as well. The DAQ could drop these events with no detriment to the resulting system sensitivity because their energy is too low. Hence, the DAQ of this system \cite{Bugalho2013, 7407512} would benefit from the online discrimination of events. The use of the filter would optimize the usage of the available BW, limiting the transmission to useful data. Therefore, we propose an online discriminating solution to be implemented on on-detector electronics, to control the data volume produced by each detector module while minimizing its impact on system sensitivity and performance. Such an on-detector and on-the-fly discriminating process would be very useful for highly pixelated PET scanners, such as the EndoTOFPET-US. It would improve the detector signal to noise ratio (SNR) reducing the system BW requirements.

The EndoTOFPET-US scanner implements one of the most recent techniques for PET detectors readout. SiPMs read by application specific integrated circuits (ASIC).
PET detectors read by ASICs are particularly suitable for the application of these
online filtering techniques given the accessibility of gamma event digital information. In the on-detector electronics, ASICs process and digitize the electrical pulse produced by the SiPM upon a gamma ray interaction on the detector crystals. The field programmable gate array (FPGA) of the on-detector DAQ front-end segment retrieves the ASICs digital data. This makes gamma events energy discrimination simple at the detector level for the DAQ module reading the gamma events' information extracted by the ASICs.

\subsection{Hypothesis and objectives}\label{sec:hyp_obj}

The initial hypothesis motivating this work is that online energy discrimination at the detector level is an effective and lightweight solution for SiPM-ASIC PET systems to address the BW limitation issue present in most highly pixelated PET scanners. To prove this hypothesis, our research must fulfill the following objectives:
\begin{itemize}
	\item Confirm the potential system benefit through a study on a reference system of the energy distribution of the received events.
	\item Verify the effectiveness of the solution studying the data reduction obtained through online energy discrimination while analyzing the possible impact on the resulting system performance.
	\item Establish the discriminating module implementation cost evaluating its resources consumption.
\end{itemize}
Confirmation of the proposed hypothesis has significant implications for PET systems. Counting on a data volume control mechanism, the system's BW usage would be optimized as well as the disk use. Thus, data volume control would reduce system requirements in terms of available BW and disk size. Additionally, such a smart event discrimination mechanism benefits system sensitivity under saturation conditions, where the first events to be discarded would be those with energies out of range. Ultimately, the analysis of online data processing techniques on digital readout (SiPM-ASIC) PET systems would open new alternatives for DAQ data handling techniques if they prove to be both effective and lightweight. There is great potential for making use of the digital information available at each DAQ level.

\section{Methods}\label{sec:meth}

This section describes the system considerations to take into account during filter design. It covers the implementation details of the various versions explored for the online energy discriminating module as well as the equipment and experiments used to analyze these solutions.

\subsection{SiPM-ASIC considerations for energy discrimination}\label{sec:consid}

\begin{figure}[htb]
	\centering
	\includegraphics[height=5cm]{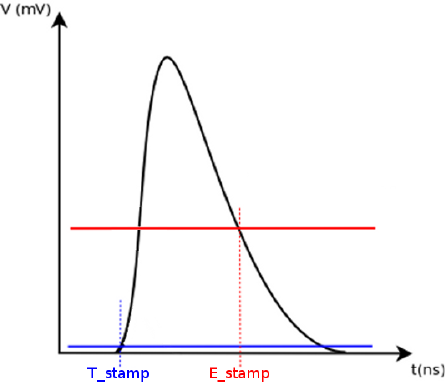}
	\caption{\label{fig:tot} Dual-threshold scheme implemented by the detector readout ASIC for TOT computation on a SiPM pulse. A low-threshold trigger tags the T\_stamp, providing the first timestamp for the TOT calculation. The SiPM pulse falling edge crossing the higher threshold discriminator sets the E\_stamp, providing the second timestamp from which the coarse TOT can be derived.}
\end{figure}

The TOFPET ASIC \cite{1748-0221-8-02-C02050}, present on the EndoTOFPET-US main detector, uses the time-over-threshold (TOT) technique to compute the energy of gamma events. For every detected SiPM pulse, the ASIC encapsulates the information of the gamma event in a digital packet. This information includes two timestamps used to calculate the TOT: the one associated with the instant when the rising edge of the SiPM pulse surpassed the lower threshold set by the ASIC (T\_stamp), and the one associated with the instant when its falling edge falls below the upper threshold (E\_stamp). The scheme shown in Figure~\ref{fig:tot} illustrates the threshold positioning and the resulting timestamps when it is applied to SiPM pulses.

Each of the two stamps contains two time components: coarse and fine time measurement. The coarse component has a 6.25 ns time binning and the fine component has a time binning of 50 ps. When we use the fine time information to compute the TOT, we achieve higher accuracy in the measure of the event's energy in comparison to using just the coarse time information. However, we considered that the precision given by the coarse time information is enough for a first level online energy discrimination. For this reason, the filter implementations analyzed in this study ignore the fine time information. We perform energy discrimination based solely on the coarse time information, thus simplifying the online discrimination modules. In practice, the online energy filter obtains each event's coarse TOT by subtracting the event coarse timestamp component associated with the T\_stamp (T\_coarse) from its coarse timestamp component associated with the E\_stamp (E\_coarse). Equation~\eqref{eq:gen_coarse_tot} defines the general expression to compute the event coarse TOT value. During online event discrimination, this coarse TOT value is compared to the defined energy window limits for decision making.

\begin{equation}\label{eq:gen_coarse_tot}
	\begin{split}
		&event\_coarse\_TOT = \\
		= even&t\_E\_coarse - event\_T\_coarse
	\end{split}
\end{equation}

\subsection{Equipment}\label{sec:equip}

Given the status of the EndoTOFPET-US system at the time when this study took place, which was pending for commissioning and calibration, we decided to test the filter performance on a similar but fully functional system, the PETsys Electronics demonstrator \cite{1748-0221-11-12-P12003}. This PET system uses the same readout electronics as the EndoTOFPET-US scanner, but the system is arranged according to a different geometry. The demonstrator integrates 2048 detector channels disposed on a small ring of detectors. Despite the differences between the two systems, all the influential characteristics for the energy discrimination process (such as ASICs, readout electronics, SIPMs, and LYSO crystals) are the same for both scanners. Thus, the conclusions obtained through the analysis of these tests on the demonstrator are valid for the EndoTOFPET-US.

\subsection{Detector energy spectrum study}\label{sec:E_sp_study}
\begin{figure}[htb]
	\centering
	\includegraphics[width=.475\textwidth]{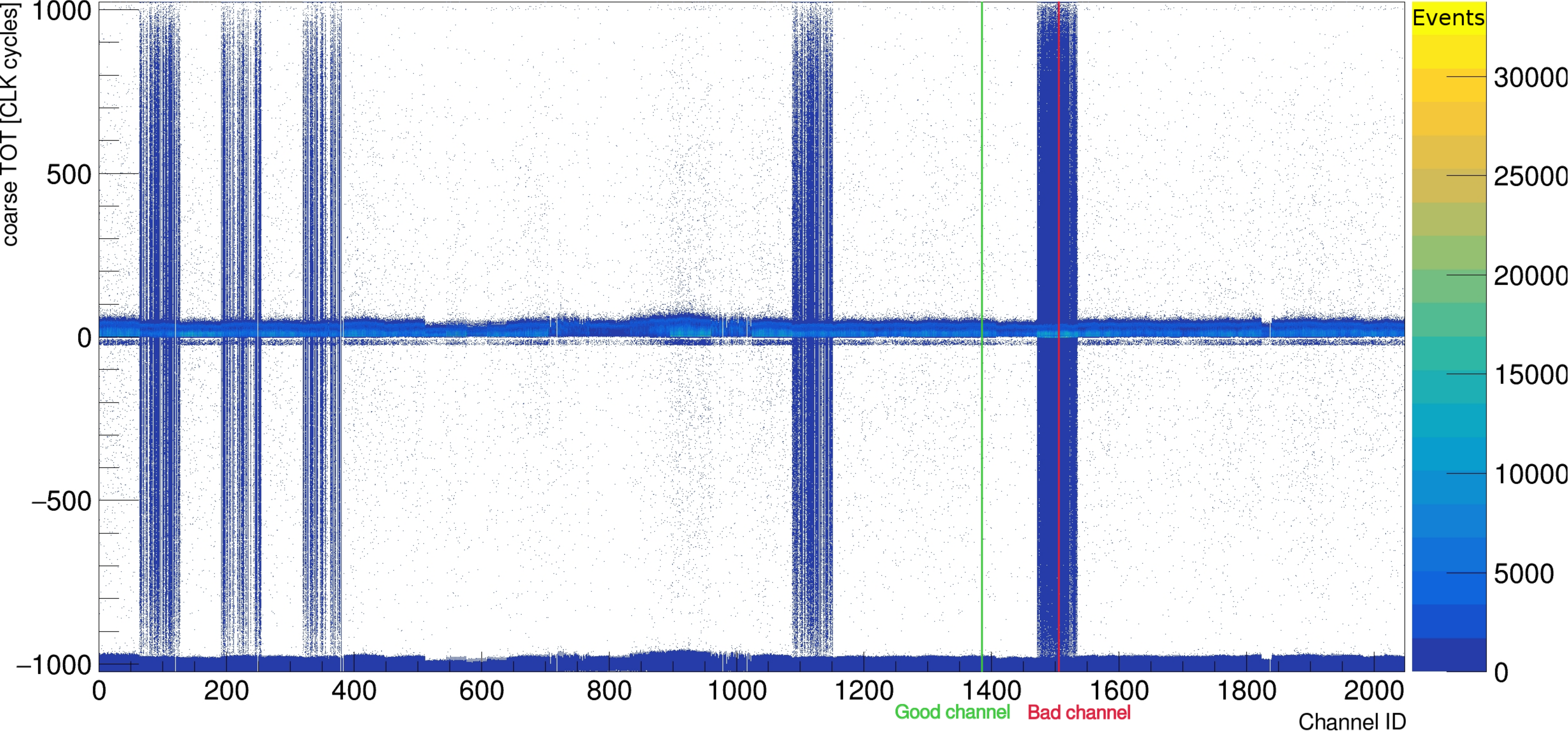}
	\includegraphics[width=.475\textwidth]{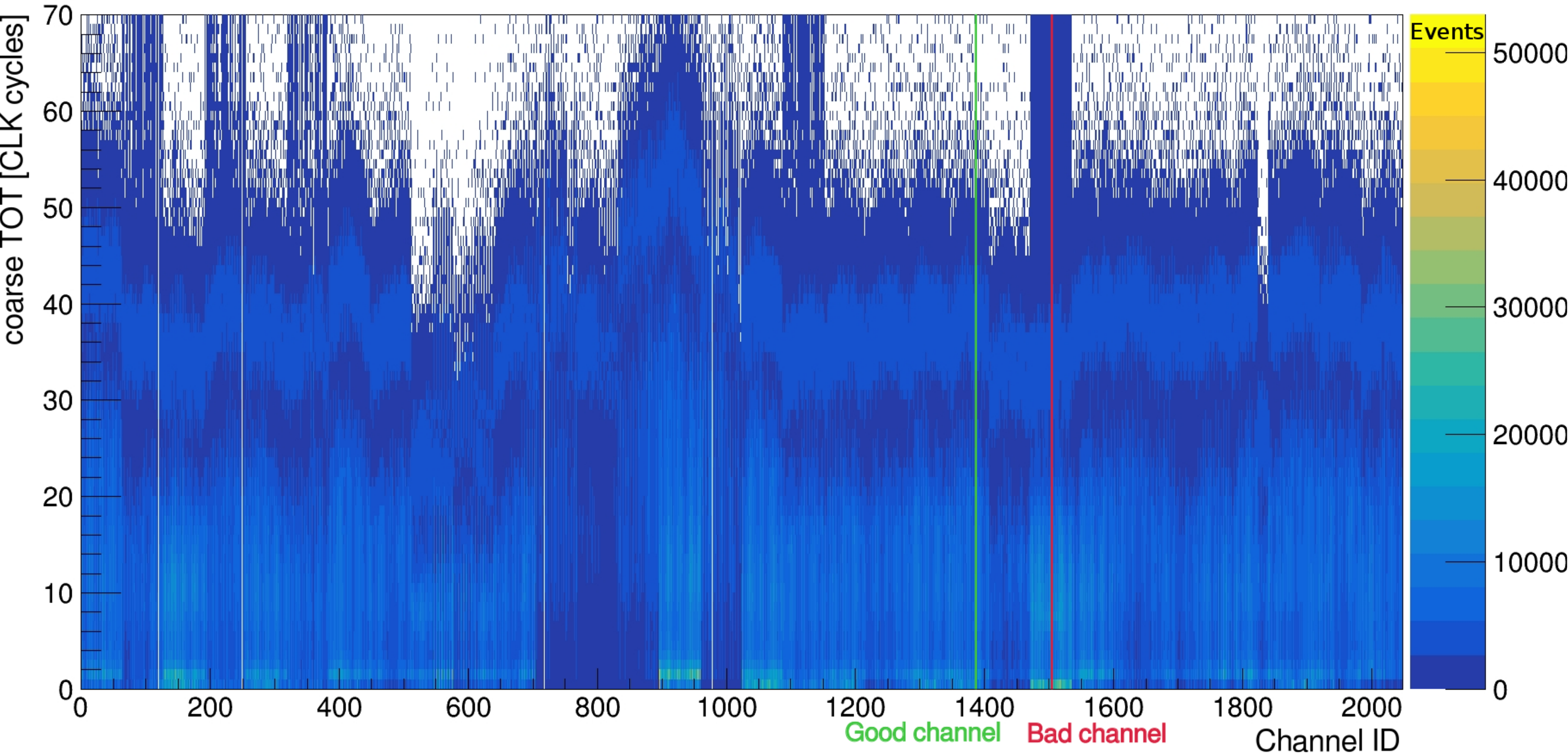}
	\caption{\label{fig:e_dist} Analysis of TOF-PET scanner energy response. Coarse TOT distribution of the received ASICs events during a 5-min acquisition with a Na22 point source. The upper plot shows the complete coarse TOT span, and the lower plot focuses on the region of interest corresponding to the valid energy range. Both plots identify two reference channels used in our study: the ``good channel'' and the ``bad channel''. These channels correspond to the system channels with the smallest and the largest amount of noise respectively.}
\end{figure}

After the first acquisition making use of the demonstrator system and through
the study of the energy space of the acquired gamma events, we noticed that the
TOFPET ASIC produces some events with incorrect timestamps, which lead to a coarse
TOT out of the expected range. The image coarse TOT map shown in the upper part of Figure~\ref{fig:e_dist} shows the energy distribution of all the gamma events obtained during a 5-min acquisition with a Sodium-22 (Na22) point source centered inside the detector ring. This graph indicates the malfunctioning of 5 ASICs on the system.
These are the ASICs with the following channel IDs: 64-127, 192-255, 320-383, 1088-1151, and 1472-1535. These five noisy ASICs produce a large number of events scattered over the whole energy space, whereas the other ASICs rarely produce events outside the valid energy range. Analyzing a detailed view of the region of interest (coarse TOT between 0 and 70 CLK cycles) depicted in the lower plot of the coarse TOT map in Figure~\ref{fig:e_dist}, we noticed that, despite being faulty ASICs, they still produce a larger number of events in this region where their energies follow the same distribution in comparison to that exhibited by the non faulty ASIC channels. This means that although they were identified as malfunctioning ASICs, these ASICs produce valid data mixed with incorrect timestamps events.

This data visualization indicates that online energy filtering would optimize a system's BW usage. The energy distribution of  events suggests that the main contribution to the non-useful data reduction would come from the exclusion of the high concentration of low-energy events considered as non-valid events. Additionally, the system would further reduce the output data rate by the discrimination of all the events cataloged as events with incorrect energy information mainly given by the malfunctioning ASICs.

The full-range coarse TOT map shown in the upper part of Figure~\ref{fig:e_dist}
conveys other important information too. The accumulation of events in the
negative coarse TOT region close to zero indicates that, for some events,
E\_coarse is slightly smaller than T\_coarse. According to the working principle
of the ASIC, this scenario should not arise. However, small length differences between the TOFPET ASIC electrical routes can cause deviations from the correct TOT. Such discrepancies from the real event energy can generate events with a negative coarse TOT close to zero for the events with the smallest energies.

Furthermore, there is another abnormal concentration of events. Some events build-up in the negative coarse TOT zone close to -1024 CLK cycles, namely, E\_coarse close to zero and T\_coarse close to its maximum value. This occurs due to what we call the counter rollover effect. Internally, the TOFPET ASIC uses a 10-bit free-running counter for the time tagging of E\_coarse and T\_coarse. If T\_coarse gets a value from the counter when it is about to roll back to its initial value (i.e. T\_coarse close to 1024), then E\_coarse would get a value shortly after rolling back (i.e. E\_coarse close to 0).

\subsection{Setting detector-specific filtering rules}\label{sec:filt_rules}
Correct identification and preservation of events with valid but misleading energy information is the key to preserving system sensitivity. Thus, the filter needs to exclude the events with a non-valid energy value while keeping the events which, despite containing meaningful energy information, lead to a coarse TOT computation laying outside the valid range due to the TOFPET ASIC particularities described in the energy spectrum study presented in Section~\ref{sec:E_sp_study}.

The acquisition software (SW) deals with incorrect energy stamps and, for their processing, it takes into account the ASIC characteristics. This SW processes received data discarding all the events that present negative energy values close to zero due to ASIC inter-path length differences. We discard these events because they present an energy close or equal to zero; hence, they are considered noise. Event-preprocessing SW also handles events suffering the counter rollover effect on their time tagging. The SW recomputes the rollover event coarse TOT by adding 1024 CLK cycles to the default computation~\eqref{eq:gen_coarse_tot}. Given the 10-bit size of the coarse time counter, this operation corrects the error introduced by the rollover effect, obtaining the appropriate coarse TOT of these events. The preprocessing SW uses a conservative energy boundary value to differentiate events undergoing the counter rollover effect. The limit selection is not adapted to the detector response but instead minimizes the possibility of filtering counter rollover events for most detector conditions. The filter implementation replicates these preprocessing SW rules, adapting and extending them to fit the detector response analyzed in the energy spectrum study presented in Section~\ref{sec:E_sp_study}.

\begin{figure}[htb]
	\centering
	\includegraphics[width=.475\textwidth]{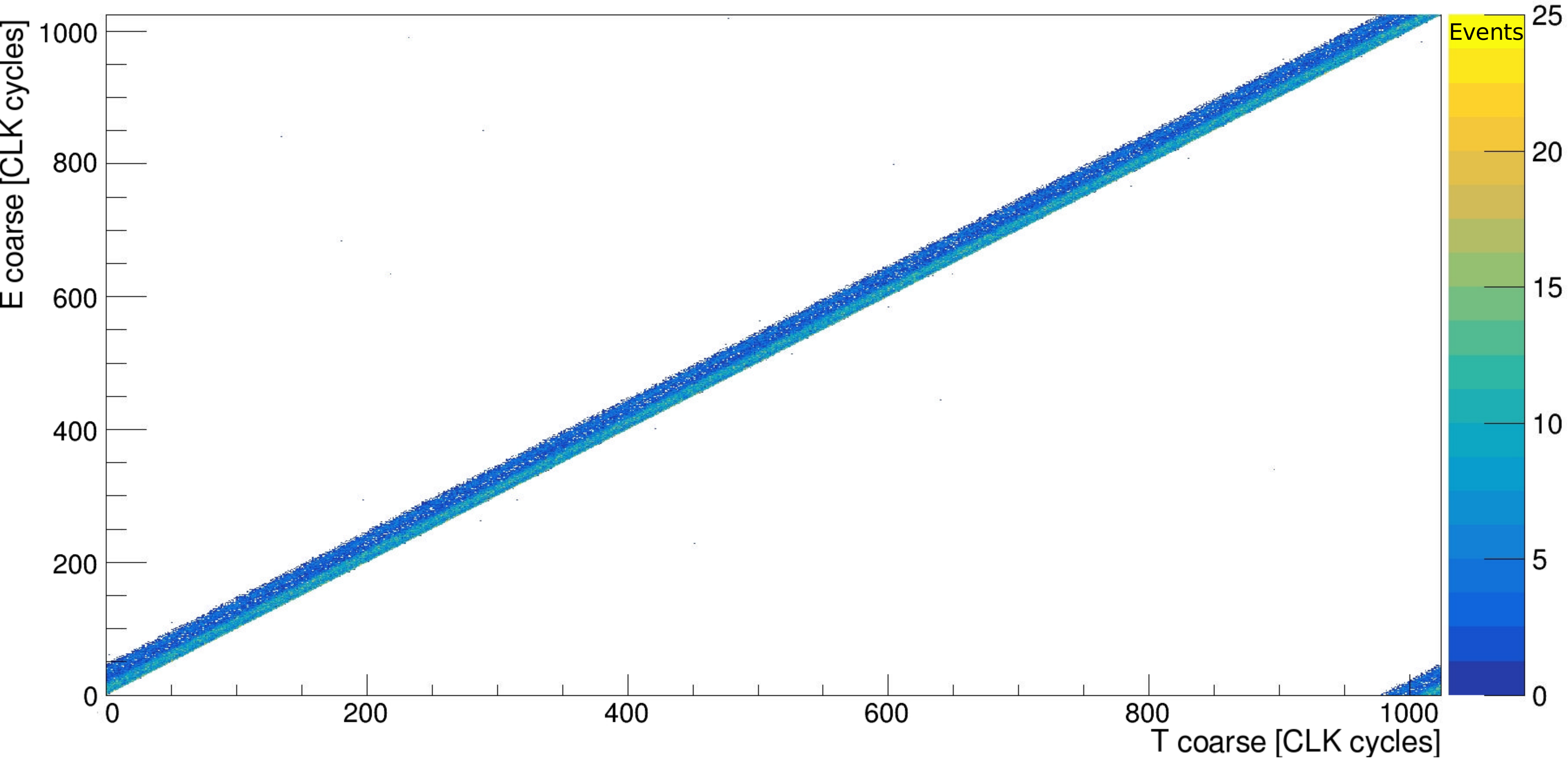}
	\includegraphics[width=.475\textwidth]{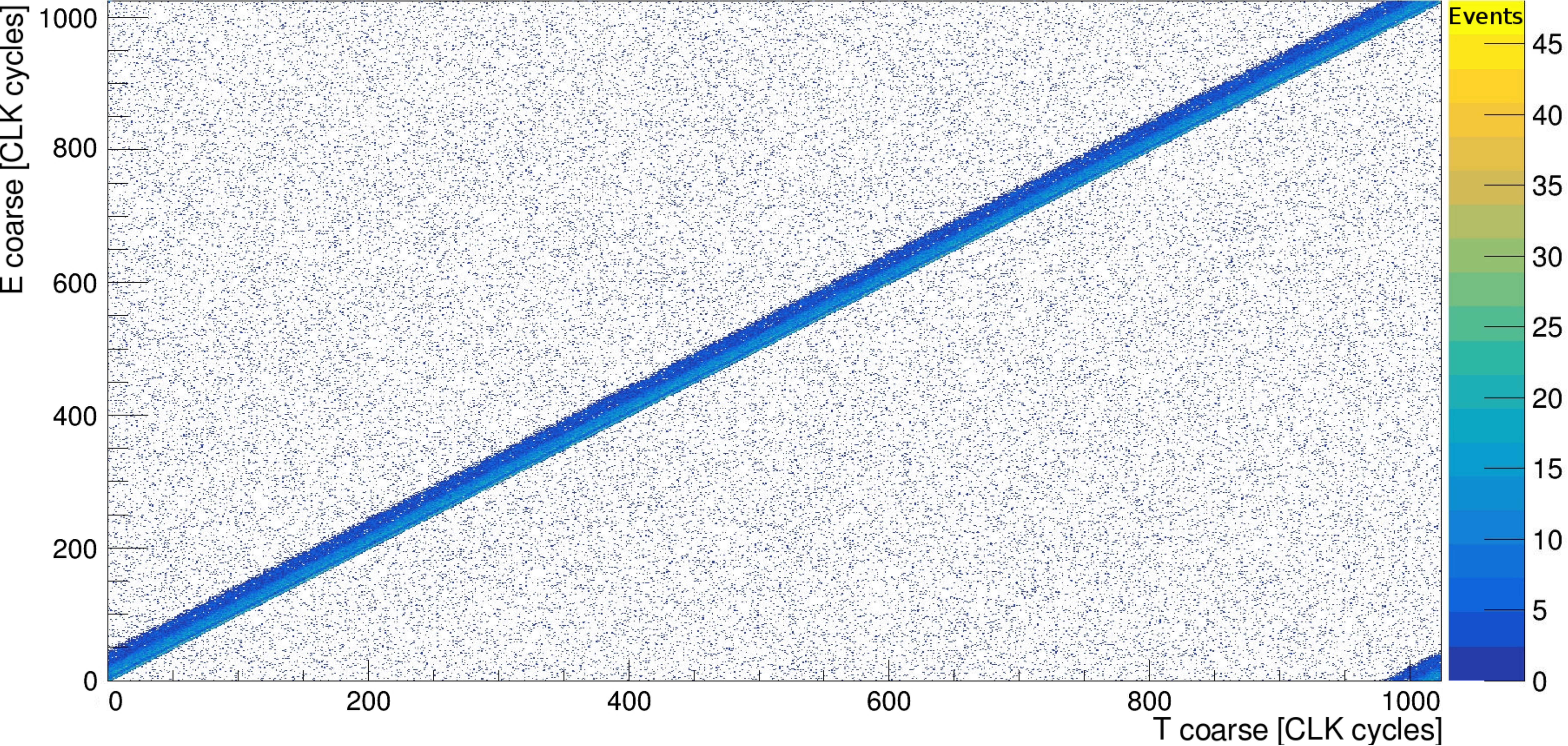}
	\caption{\label{fig:ref_ch_tot_distr} Analysis of TOF-PET scanner energy response. E\_coarse versus T\_coarse distribution of the received ASICs events expressed in CLK cycles with 6.25 ns steps. Color scale indicates the number of events. The produced events are the result of a 5-min acquisition placing a Na22 point source centered on the detector ring. The upper plot shows distribution for the reference ``good channel'' and the lower plot shows the distribution for the ``bad channel''.}
\end{figure}

To visualize the benefit of setting detector-specific filtering rules we make a visual comparison of the acquired data energy distribution for the two reference channels highlighted in Figure~\ref{fig:e_dist}, the ``good channel'' and the ``bad channel''. We selected the ``bad channel'' as the system channel producing the largest number of events; thus, it would be the noisiest system channel. Channel 1506 is the reference ``bad channel''. In opposition, the reference ``good channel'' is the one producing the smallest number of events, the least noisy channel (i.e. the one with fewest events with an energy outside the valid energy range). We chose channel 1390 for this purpose. To analyze the energy distribution on a reference channel we plot the T\_coarse versus E\_coarse map of the events it produces. The timestamp distribution plots in Figure~\ref{fig:ref_ch_tot_distr} show the resulting reference maps for the ``good channel'' in the upper graph and for the ``bad channel'' in the lower graph. By setting detector-specific filtering rules, the events cataloged as noise can be eliminated. As a consequence, the newly obtained T\_coarse versus E\_coarse plots of the ``bad channel'' should be similar to the reference ``good channel'' map after detector-specific filtering rules are set.

\subsubsection{Sensitivity protection noise rejection frontier}\label{sec:ro_search}
To select the optimum coarse TOT limit value that keeps the counter rollover events  while rejecting other non-valid events with negative coarse TOT values, we studied the coarse TOT distribution on the channel producing the highest average TOT values. We designate this new reference channel as the ``highest TOT channel''. By choosing this as the reference channel, we avoid setting an upper limit that would cut part of the valid events on the channels producing higher coarse TOTs than the average.  The coarse TOT
distribution of Figure~\ref{fig:e_dist} illustrates how the ASIC with channel IDs [896:956] generates events with higher coarse TOT values. Out of these channels, 925 is the one producing the highest coarse TOTs. Thus, channel 925 is the one selected for reference.

\begin{figure}[htb]
	\centering
	\includegraphics[width=.475\textwidth]{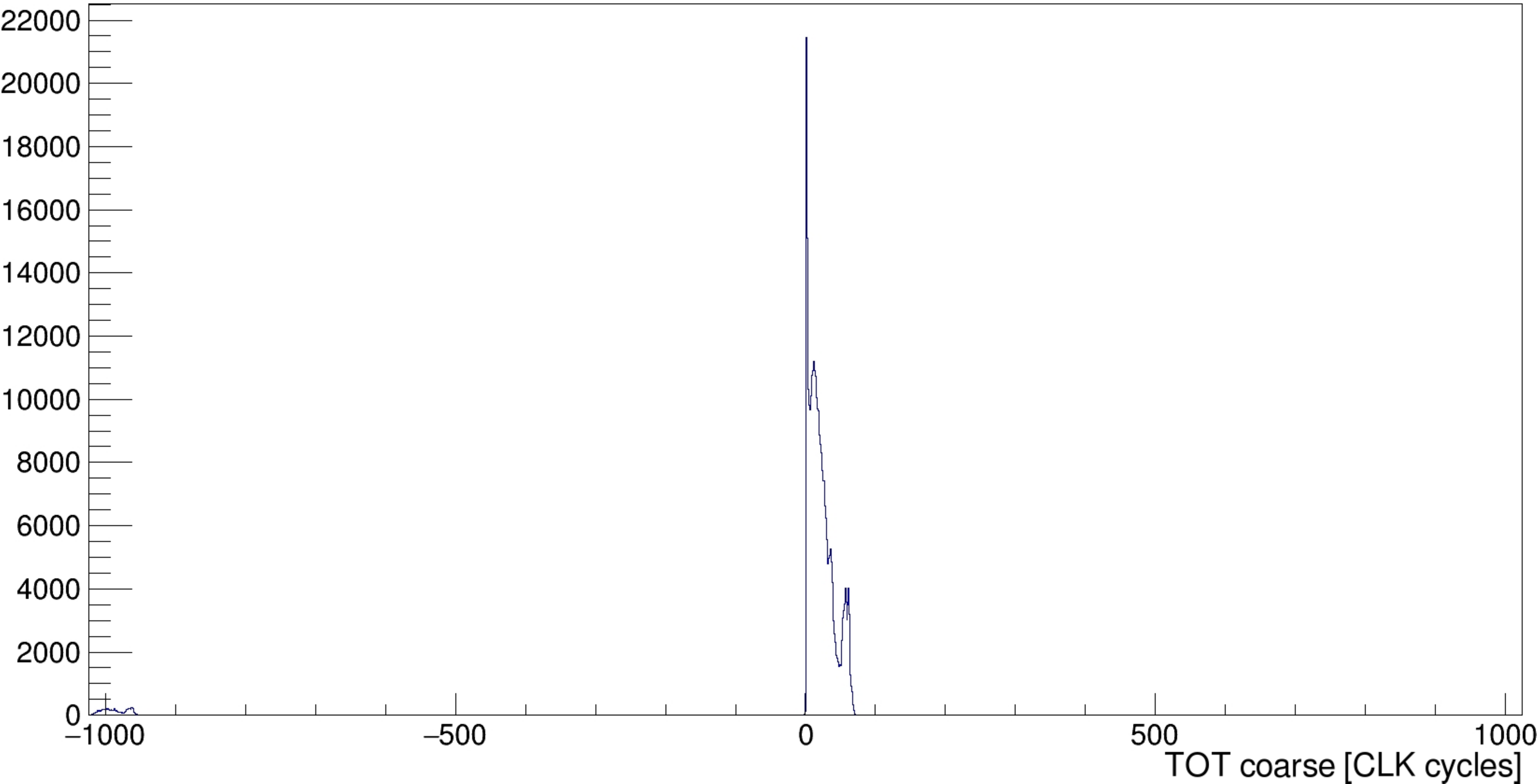}
	\includegraphics[width=.475\textwidth]{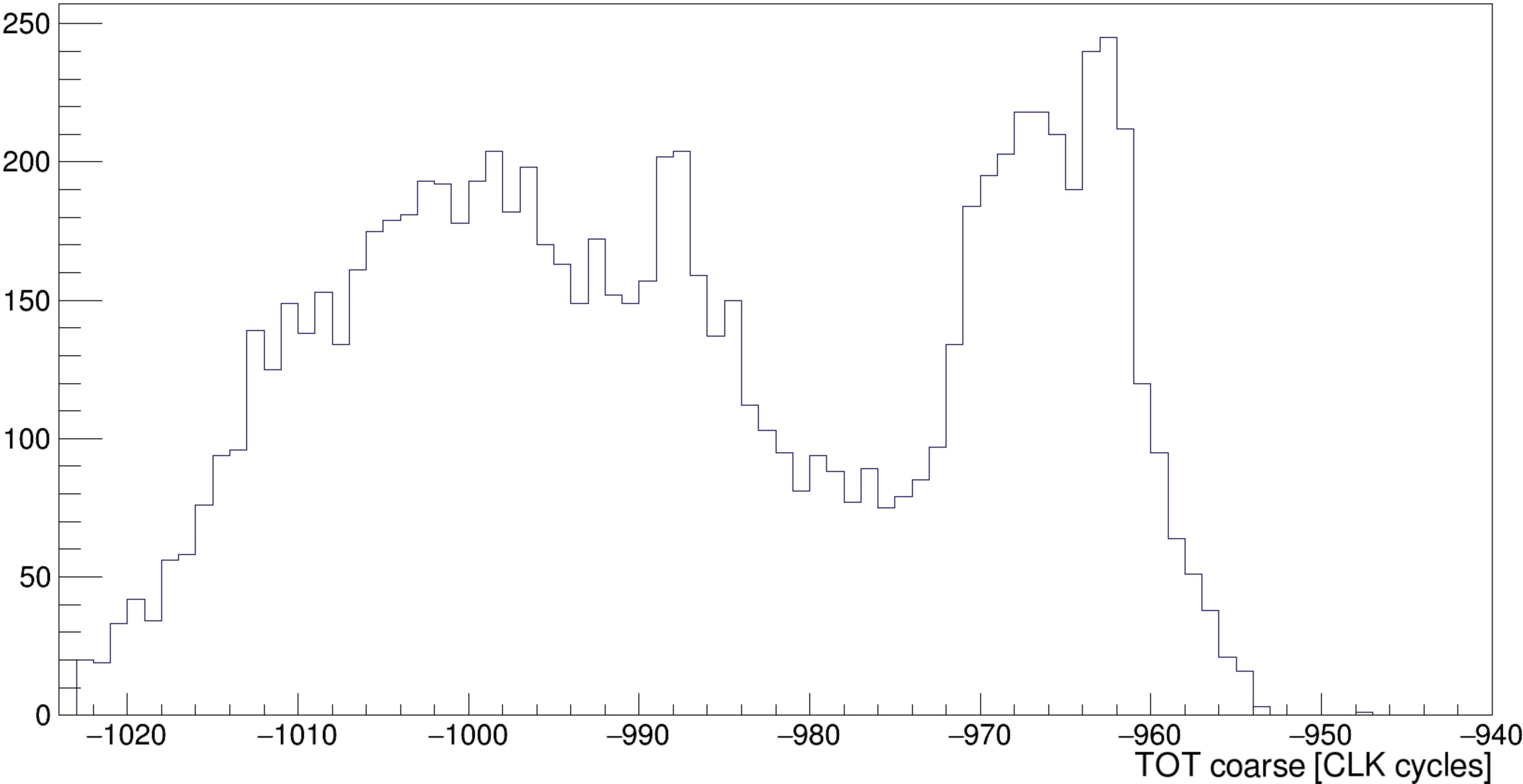}
	\caption{\label{fig:coarse_tot_dist} Analysis of TOF-PET scanner energy response. Coarse TOT histogram produced by the received ASIC events for the reference system channel producing the highest coarse TOT values. We express coarse TOT  in CLK cycles with 6.25 ns steps. The produced events are the result of a 5-min acquisition placing a Na22 point source centered on the detector ring. The upper plot shows the complete coarse TOT span and the lower plot focuses on the TOT region concentrating the events that suffered the counter rollover effect.}
\end{figure}

The full-span coarse TOT distribution in the upper plot of Figure~\ref{fig:coarse_tot_dist} illustrates how, for the reference channel, most of the events lay in the region \(coarse\_TOT = [0,80]\) CLK cycles. Events in this region do not undergo any of the TOFPET ASIC particularities. This histogram also shows a smaller concentration of events in the area \(coarse\_TOT = [-1024,-950]\); these are rollover events. Observing the distribution of events in this region, as shown in the lower plot of Figure~\ref{fig:coarse_tot_dist}, we extracted the maximum coarse TOT value found in this region to define the counter rollover limit. We obtained the limit value to set on the filter by adding a 3-CLK-cycle margin to the maximum coarse TOT value of the five acquisitions. As a result, we set the counter rollover coarse TOT limit to -950 CLK cycles. Adding a wider margin of CLK cycles would imply the filtering of fewer noise events, but adding a smaller one could cause the filtering of meaningful counter rollover events. We chose 3 CLK cycles to balance these scenarios.

Using this information, we altered the general coarse TOT computation~\eqref{eq:gen_coarse_tot} for the TOFPET ASIC special cases obtaining the coarse TOT rollover case equation, equation~\eqref{eq:ro_coarse_tot}, and the coarse TOT path length differences case equation, equation~\eqref{eq:pd_coarse_tot}. These filtering rules are hardwired in the energy filter implementation.

\begin{equation}\label{eq:ro_coarse_tot}
	\begin{split}
		co&arse\_TOT= 1024 + E\_coarse - T\_coarse \\
		&\iff (E\_coarse - T\_coarse) \leq -950,
	\end{split}
\end{equation}

\begin{equation}\label{eq:pd_coarse_tot}
	\begin{split}
		&coarse\_TOT= 0 \iff \\
		-950 < &(E\_coarse -  T\_coarse) < 0.
	\end{split}
\end{equation}

Using these new filtering rules we act on the \(E\_coarse < T\_coarse\) region, discarding noise events while preventing rejection of valid events undergoing the counter rollover effect. However, in light of the observed ``bad channel'' energy distribution shown in Figure~\ref{fig:ref_ch_tot_distr}, we conclude that setting a filtering rule to act on the \(E\_coarse > T\_coarse\) region will further reduce the volume of data produced by a noisy ASIC channel.

\subsubsection{Detector upper energy bounds}\label{sec:max_search}
\begin{figure}[htb]
	\centering
	\includegraphics[width=.475\textwidth]{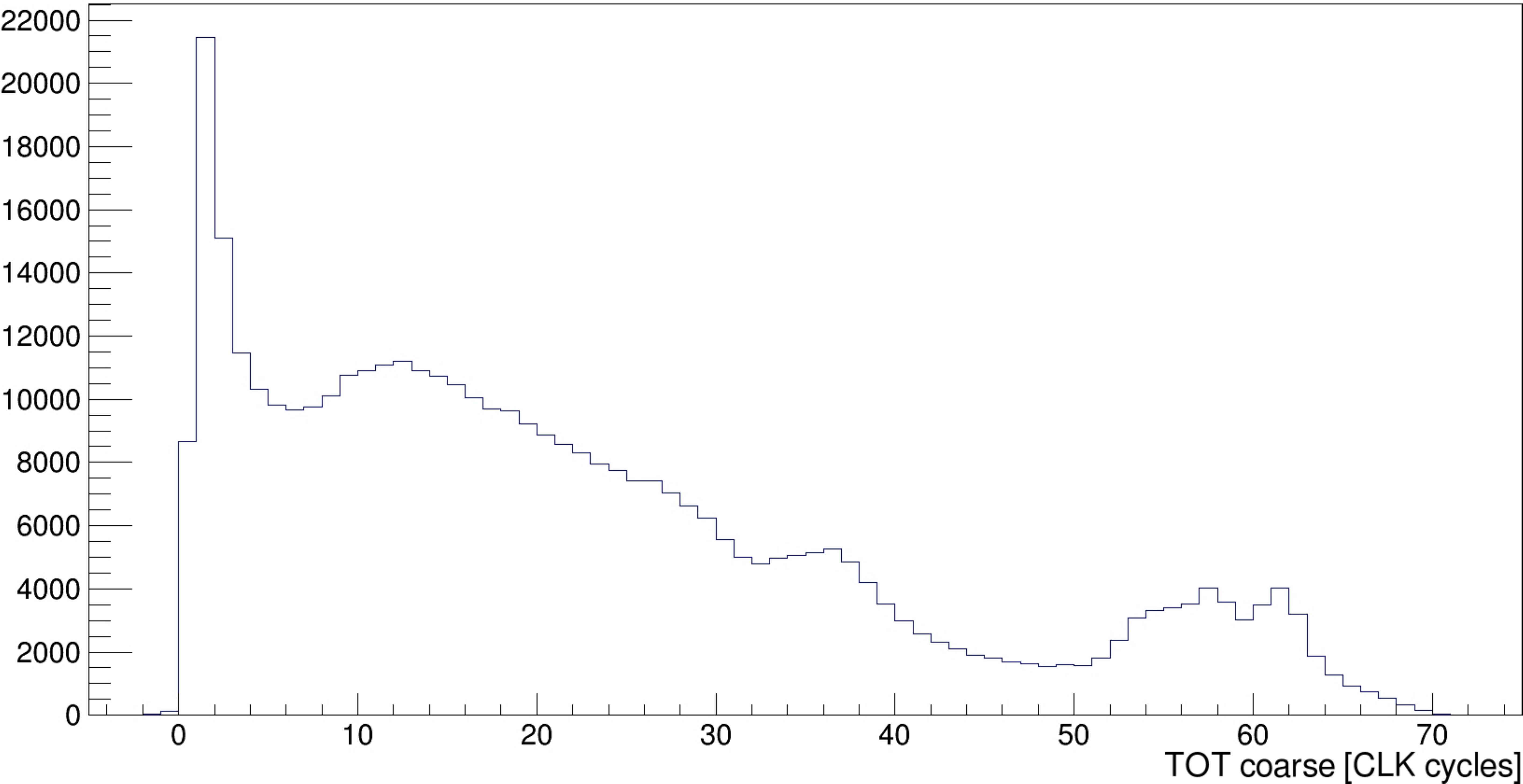}
	\caption{\label{fig:totH_ROI} Analysis of TOF-PET scanner energy response. Coarse TOT histogram produced by the received ASIC events for the system's reference channel ``highest TOT channel'' expressed in CLK cycles with 6.25 ns steps. The produced events are the result of a 5-min acquisition placing a Na22 point source centered on the detector ring. The plot focuses on the valid energy range.}
\end{figure}

To eliminate noise events from the \(E\_coarse > T\_coarse\) region, we had to determine a fixed upper limit on the event acceptance window. The methodology followed to select an appropriate limit imitates the one used for the counter rollover boundary choice presented in Section~\ref{sec:ro_search}. For noise rejection in the positive energy space, we analyzed the energy distribution from the reference ``highest TOT channel'' within the region \(coarse\_TOT = [0,80]\) as shown in the coarse TOT plot of Figure~\ref{fig:totH_ROI}. To obtain the upper-bound value, we studied the histograms obtained and extracted the maximum recorded value adding a 3-CLK margin. We selected this margin according to the same criteria as those used for the rollover boundary search: balance of noise reduction versus data loss.

The online energy filter sets the coarse TOT of the events surpassing the upper limit to zero, as shown in the new coarse TOT equation for high energies,
equation~\eqref{eq:ub_coarse_tot}. The DAQ filtering stage deletes all events with a zero coarse TOT from the data packet, including those exceeding the upper limit or the rollover boundary, when any low threshold setting different from zero is used. This upper energy bound is hardwired in the energy filter implementation along with the filtering rules expressed by equations~\eqref{eq:ro_coarse_tot} and~\eqref{eq:pd_coarse_tot}.

\begin{equation}\label{eq:ub_coarse_tot}
	\begin{split}
		&coarse\_TOT= 0 \iff \\
		(E\_&coarse -  T\_coarse) \geq 80
	\end{split}
\end{equation}

\subsubsection{Filtering rules effect}\label{sec:rules_check}
\begin{figure}[htb]
	\centering
	\includegraphics[width=.475\textwidth]{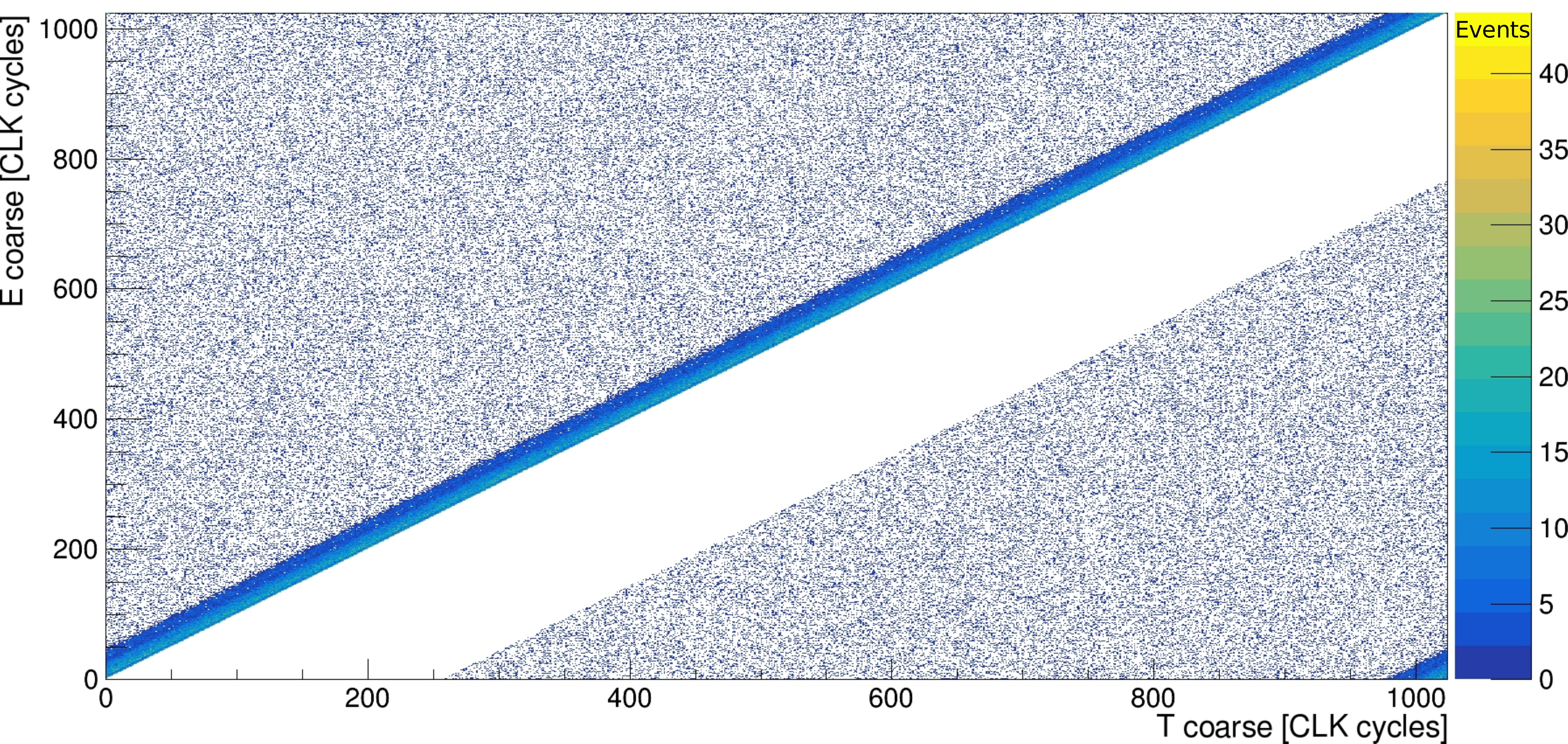}
	\includegraphics[width=.475\textwidth]{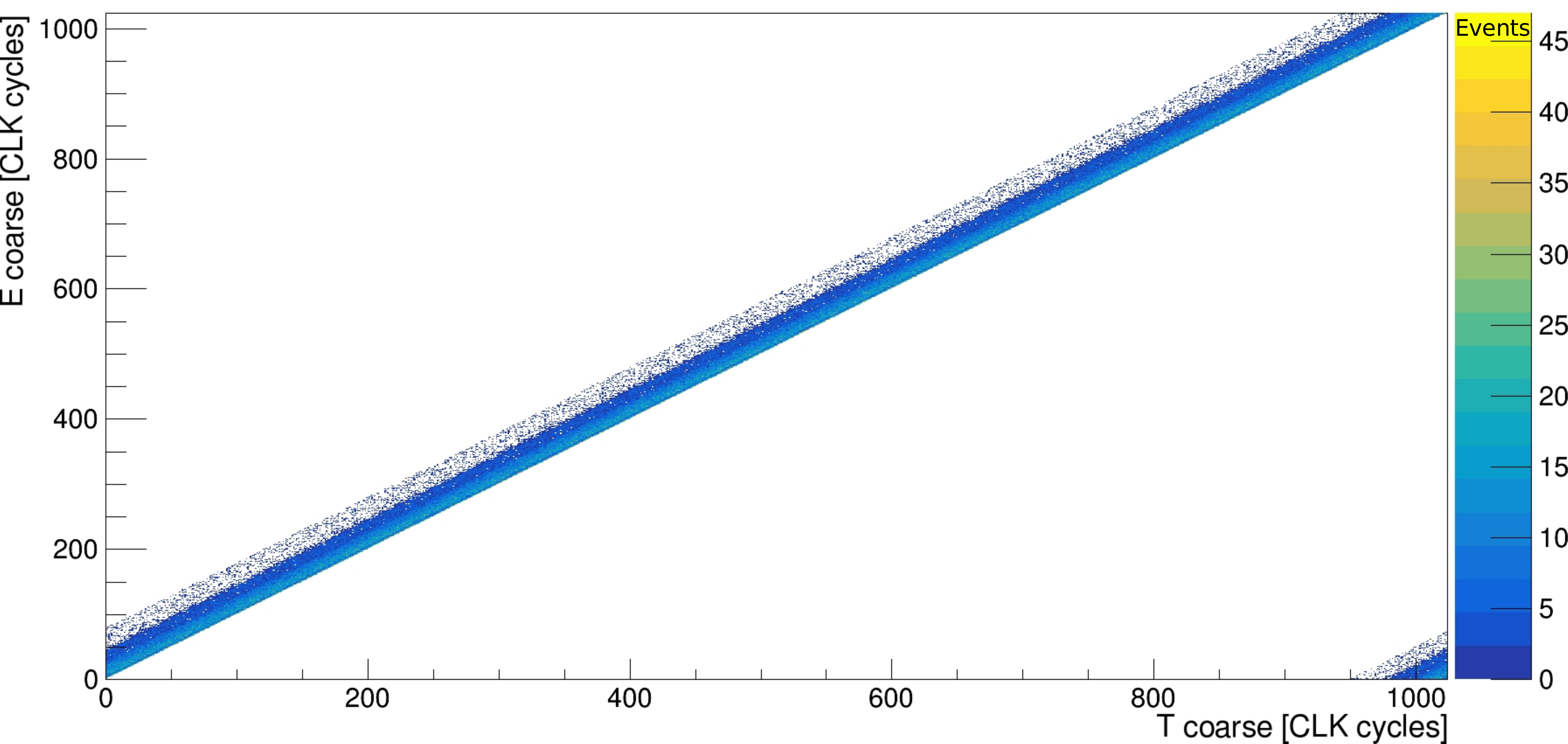}
	\caption{\label{fig:rules_comp} Analysis of TOF-PET scanner energy response. E\_coarse versus T\_coarse distribution of the received ASIC events produced on the reference ``bad channel'' expressed in CLK cycles with 6.25 ns steps. The received events are the result of a 5-min acquisition placing a Na22 point source centered on the detector ring. The upper figure shows the case when the online discriminator implements the general filtering rules defined on the preprocessing SW. The lower plot represents the detector specific rules case.}
\end{figure}

For a quick assessment of the noise reduction achieved through the new filtering rules we make use of the visual comparison method introduced in Section~\ref{sec:filt_rules}. Thus, we can verify the correct adaptation of the initial filtering conditions derived from the preprocessing SW to the observed system response. The E\_coarse versus T\_coarse distribution graphs presented in Figure~\ref{fig:rules_comp} show the effect of applying the general preprocessing SW rules (upper graph) and the detector-specific rules (lower graph) on the online energy filter.

Considering the data acquired making use of the online energy filter we can deduce that the general rules ensure preservation of events suffering the counter rollover effect, while they reject some of the non-valid events with negative coarse TOT values. However, much more noise can be eliminated by setting detector-specific rules. As seen in the lower timestamp distribution map in Figure~\ref{fig:rules_comp}, the new detector-specific rules eliminate all the noise that can be cleared using the same unique fix margin for every channel while preserving all relevant events, even for the channels producing the highest TOT values.

\subsection{Filter implementations}\label{sec:imp}

Since the SIPM pixels and ASIC channels of the system do not all achieve exactly the same performance, the response of the full system is non-uniform. The solution to this problem lays in the calibration process \cite{7407512}. For each system channel, an energy calibration procedure obtains a table mapping the measured TOT and the corresponding gamma event energy in keV units. The data processing SW uses this information to obtain the actual energy measurement for each gamma event. Regarding firmware (FW) implementation of the filter contained in the front-end electronics FPGAs, we decided to start by designing an uncalibrated version of the filter followed by its calibrated counterpart. The uncalibrated implementation ignores the energy calibration information setting for all the channels the same TOT thresholds, which define the energy acceptance window. However, the calibrated version reads the energy calibration file to set channel-specific energy thresholds for every single detector channel. This approach enabled us to begin with a quick proof of concept using the simpler and faster to implement uncalibrated filter version. Once we finished the first stage of testing and tuning, we focused our efforts during the implementation of the calibrated filter version on calibration-related issues.

\subsubsection{Uncalibrated energy filter}\label{sec:uncal}

Complete filter implementation consists of two FW modules for filter configuration and event discrimination, plus a Python SW script. The Python SW script makes use of the PETsys Electronics SW libraries to configure and initialize ASICs and DAQ front-end modules. Then the script sends a filter configuration command to every active DAQ front-end module to set the upper and lower thresholds for the energy acceptance window of the DAQ module's filter. After system and filter configuration is completed, the script reads data through the DAQ.

The TOT-energy relationship varies from channel to channel. Thus, without a system transversal energy to TOT conversion, the uncalibrated filter must define its thresholds in terms of coarse TOT units, and so does the SW script for filter configuration. Without the information of the various channel responses, the uncalibrated filter uses a single pair of thresholds for the entire system. The discriminating stage accepts only those events whose coarse TOT is between the temporal thresholds set for the filter independently of the actual energy that it represents for the channel under consideration. Even though the user specifies the filtering limits in seconds, ASICs coarse time tags are expressed in CLK cycles. Hence, to build filter configuration commands, the SW script converts the TOT filtering limits to CLK cycles, avoiding unnecessary operations on FW, where these values are directly compared against the event coarse TOT.

During the detector configuration phase, the filter configuration module receives the threshold setting command and stores the module's filter thresholds in a register to be read by the data filtering module. On the other hand, online discrimination implies continuous inspection of incoming ASIC data packets. For every gamma event contained in a data packet, the filter computes the event coarse TOT either using the  general case equation~\eqref{eq:gen_coarse_tot} or one of the special case equations~\eqref{eq:ro_coarse_tot},~\eqref{eq:pd_coarse_tot} or~\eqref{eq:ub_coarse_tot}. Then the filter compares the resulting event coarse TOT against the given energy thresholds and makes a filtering decision. The filter removes from the data packet the events rejected by energy, sending to the DAQ collector module packets containing only the valid gamma events of the original data packet.

Inside the ASIC, data is handled as 8-bit words, implementing an 8-bit architecture,
and it sends data packets with all the detected gamma events every \(6.4\ \mu s\), that is, every 1024 CLK cycles. Internally, the DAQ system formats incoming ASIC data packets to deal with 64-bit words: it implements a 64-bit architecture. Each 64-bit word contains either a packet header word or a complete gamma event, and the FW handles each word as a whole. As a consequence, the DAQ front-end module compacts the data packets produced by the ASIC as a continuous data stream. Thus, in the front-end FPGA, the FW processes the incoming data stream in a non-continuous burst mode. Setting a 64-bit architecture on the DAQ reading an 8-bit architecture ASIC, the DAQ dead time probability is reduced. The uncalibrated filtering module needs one CLK cycle to process each data packet word, but complete processing of a data packet takes one additional CLK cycle. Given the DAQ non-continuous packet management, in which data packet processing takes a maximum of 128 CLK cycles, the online energy filter module has at its disposal 896 free CLK cycles. Thus, one extra CLK cycle for the data packet processing do not cause a problematic delay.

The DAQ on-detector front-end electronics reading the ASICs contain two specific FW modules for online energy filtering, namely, the filter configuration and data discrimination modules. While the DAQ off-detector collector module needs no adaptation because the filter functionality is transparent to it. The filter configuration commands use the common front-end  format for configuration and control commands. Thus, the DAQ collector module, which also handles the front-end control and configuration,
merely retransmits these commands to the addressed front-end unit. From
the point of view of data discrimination, the filter simply removes events that are not in compliance with the energy rules without modifying the data packet format. Thus, the DAQ collector module  needs no modifications; it processes filtered data packets as regular data packets.

The filter does not alter event information, it just reads the event
coarse timestamps to compute its coarse TOT value for internal comparison
to the energy window limits. Events whose energy meets the acceptance
conditions pass the online filter without being modified. When the filter module
identifies an event with an incorrect energy value due to one of the two ASIC special
cases, coarse timestamps could be modified on the fly to compensate the effect of these cases. Nonetheless, we decided to leave the stamps unmodified on the first filter implementation to avoid introducing any artifact.

\subsubsection{Calibrated energy filter}\label{sec:cal}
Given the differences, between the detector channels' responses, the system
requires a calibration process to homogenize the response of the 2048 channels.
This detector calibration process includes energy calibration, which relates TOT and its equivalent energy for every system channel. The calibrated filter version exploits the energy calibration information to set channel-specific filtering thresholds. As a consequence, calibrated filter thresholds can be defined as real energy values expressed in electronvolt [eV].

The TOT measurement of SiPM pulses is strongly nonlinear \cite{ORITA2011S24}, and we cannot model the TOT-energy relationship analytically. Thus, the calibration process obtains this relation experimentally by measuring the resulting TOT from gamma photon sources with photo-peaks of known energy. Our experimental data shows an exponential TOT-energy relation, and most of the system channels' energy responses can be fitted with exponential functions. A conversion table is created during calibration to define each channel's exponential function fit. This table relates the TOT and energy for each channel within an energy subrange. This ranges from \(TOT=0\) to \(TOT=499.5 ns\) with 0.5 ns steps. According to the experimental results, \(TOT=499.5 ns\) corresponds to a value around 100MeV for most of the channels. This value far above our energy range of interest, i.e. around 511 keV, introduces a margin for the ASIC channels producing the highest TOTs.

We modified the filter SW script for the calibrated version to configure
independent energy thresholds for every system channel. The script receives two parameters specifying lower and upper filtering thresholds expressed as energy values in keV. These values define the filter energy cuts or acceptance window. For each system channel, the script reads the TOT-energy conversion table searching for the closest energy values to the given filter thresholds and obtains their equivalent TOT values. Channels exhibiting an uncommon response, such that the calibration algorithm in search of the best exponential function fit cannot converge, are configured with the filter thresholds of previously configured channel that presented a valid function fit. Configuration commands express threshold values as TOT limits in CLK cycles, as in the uncalibrated filter version, because the ASIC data packet uses the same units; thus, extra FW processing is avoided. The SW script repeats this procedure for each channel to configure the online energy filter thresholds for every system channel independently. After configuration, the SW script launches the acquisition through the DAQ system.

Given the configuration commands' channel specificity, command syntax changes to include a channel's address. The calibrated filter configuration command follows this syntax at the DAQ front-end level:
\begin{itemize}
	\item A header byte to identify the command type, filter threshold setting command
	\item A 10-bit word containing the channel ID
	\item A 10-bit word containing the lower threshold value in CLK cycles
	\item A 10-bit word containing the upper threshold value in CLK cycles
\end{itemize}
Although the system contains 2048 channels, the configuration command channel ID contains 10 bits instead of 11 because it must identify the channel within the on-detector front-end module, which contains 1024 channels. The DAQ off-detector segment encapsulates the filter configuration command using a header that already identifies the destination front-end module.

For the storage of threshold values, on each front-end detector module, the FW
implements a 20-bit wide look up table (LUT) with 1024 entries, one for each of its channels. This LUT is 20 bits wide to store both lower and upper thresholds. The filter configuration FW module unpacks each command, extracting its destination channel ID and thresholds. It uses the channel ID for LUT addressing and writes its threshold values. During the configuration phase, this module receives a total of 1024 configuration commands from the SW script to fill the complete front-end module LUT.

During the acquisition phase, the online energy filtering stage continuously inspects incoming ASIC data packets. This FW module extracts individual gamma event information  from the data packets to apply the following process:
\begin{enumerate}[1.]
	\item Read event channel ID and timestamps.
	\item Compute event coarse TOT using the  general case equation~\eqref{eq:gen_coarse_tot} or one of the special case equations~\eqref{eq:ro_coarse_tot},~\eqref{eq:pd_coarse_tot} or~\eqref{eq:ub_coarse_tot}.
	\item Read channel thresholds from the LUT using the event channel ID for addressing.
	\item Compare the event coarse TOT to the thresholds.
	\item Make event rejection decisions.
\end{enumerate}

Reading from the LUT adds two CLK cycles for the processing of each gamma event to the one CLK cycle required by the uncalibrated version. Complete data packet processing, as for the uncalibrated filter, takes one additional CLK cycle. Given the different architectures implemented by the ASIC and DAQ front-end modules (8-bit versus 64-bit), as explained in Section~\ref{sec:uncal}, the filter has 896 CLK cycles at its disposal for the extra processing time that it requires. Hence, the filter can handle a maximum of 447 events per data packet, 70Mevents/s, causing no problematic delay. Simulations show how such kind of electronics produce no more than 10Mevents/s per DAQ front-end module  \cite{Frisch2013}, which leads us to the conclusion that the online energy filter can absorb the volume of data produced by the detector modules on-the-fly,  incurring no dead time.

The calibrated version has no special requirements for the DAQ collector module either. The DAQ off-detector segment forwards filter configuration commands and processes incoming filtered data packets as it does for any other configuration command or data packet.

\subsection{Experiments and laboratory methods}\label{sec:exp}

Our experiments emulated a real PET acquisition scenario for online energy filter performance characterization. A Na22 point source centered in the detector ring radiated uniformly over the detector's surface. Based on source aging, we estimated that the point source used in our experiments produces an activity of around \(24\ \mu Ci\). Na22 sources are positron emitters in the same way that radioactive isotopes used for PET tracers are. Once a positron is annihilated with an electron, the process emits two back-to-back gamma photons with a 511 keV energy. Our experiments targeted the detection of these gamma rays as in real PET acquisition. Having a source illuminating the detectors with the right energy, we could verify the correct data preservation in the valid energy range, while noise produced outside this range was filtered.

The main goal of this work was to reduce the data volume excess produced by the system. Therefore, our experiments analyzed the influence of the filter on the system's
data output rate. We studied the data reduction obtained through the filter's low-energy threshold sweep for various filter implementations. All acquisitions were performed under the same conditions to ensure a common framework for comparison of the results. In the experiments a constant acquisition time was maintained, and the same radioactive source was used. By maintaining these two experimental conditions, we could ensure that the average volume of data produced by the detector modules would remain constant.

A proper data reduction process must preserve system sensitivity, which directly affects performance. To verify that the system's performance was maintained, in our final experiment, we placed the Na22 source in the center of the detector ring and performed a long enough acquisition to obtain good statistics for the analysis of the system's energy spectrum and CTR distribution. Specifically, using a \(24\ \mu Ci\) Na22 source, we found that 10-min acquisitions were required to obtain enough statistics to produce smooth and contiguous histograms.

\subsection{Statistical methods}\label{sec:stat_m}
The conclusions obtained from our tests are not based on the result of a
single acquisition per experiment. To ensure reliability of the results, we  performed several acquisitions for every experiment. From these acquisitions we extracted their average values and variabilities and used them to express our results. Given the consistency of the results, we decided to limit the number of acquisitions per test to 5.

For the statistical data analysis and its visualization we used ROOT, a framework for statistical analysis and visualization. Through this framework, we read the gamma event database created with each acquisition, extracted the event's feature or features of interest for the experiment, and visualized the data in the form of 1D or 2D histograms showing the distribution of the values obtained for the parameters under study.

\section{Results}\label{sec:res}

This section describes the results obtained making use of the two different
versions of the online energy filter described in Section~\ref{sec:meth}.

\subsection{Uncalibrated energy filter}\label{sec:res_uncal}
\begin{figure}[htb]
	\centering
	\includegraphics[width=.475\textwidth]{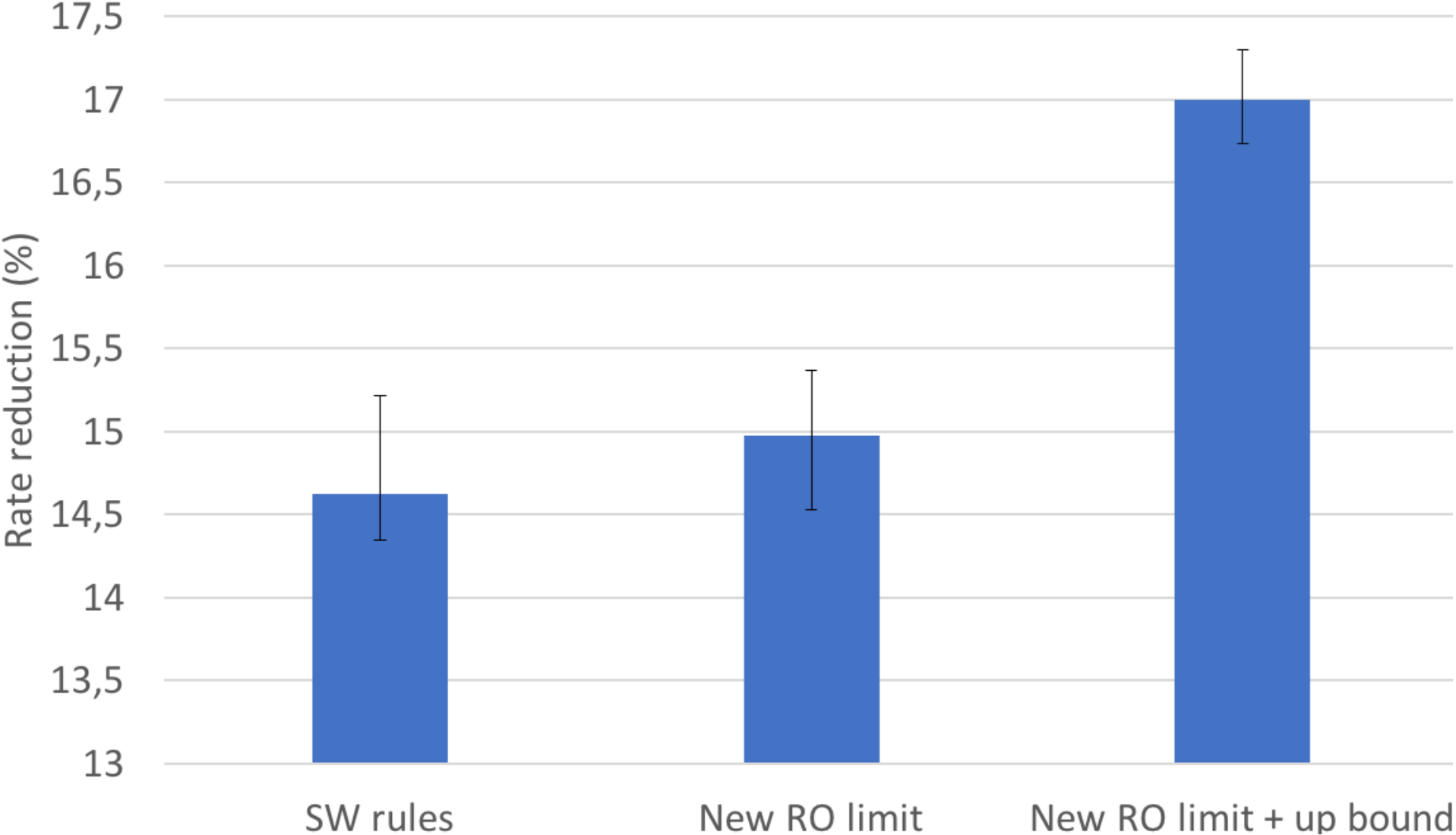}
	\caption{\label{fig:red_by_opt} Rate reduction obtained through the uncalibrated online energy discriminator using various filtering rules implemented on the TOF-PET scanner DAQ front-end modules. We express the rate reduction as a percentage relative to the reference non-filtering case.}
\end{figure}

During the definition of the filtering rules described in Section~\ref{sec:filt_rules}, we implemented various intermediate filter versions. Setting no thresholds on these filter versions and measuring the obtained data rate would give us the contribution of each filtering rule to the non-valid data rate reduction. To obtain the reference volume of data for comparison, we ran an additional set of acquisitions under the same experimental conditions but using no filter. The comparative column chart in Figure~\ref{fig:red_by_opt} shows the data reduction obtained through the various filtering rules. For each implementation we computed the average value of the 5 acquisitions and compared them to the reference non-filtering case to get the
percentages of the data volume reduction. The error bars show the variability
of the data volume for acquisitions under these circumstances.

Setting no filter thresholds, this experiment gives us the raw contribution of
each hardwired filtering rule. As seen in Figure~\ref{fig:red_by_opt}, with the  filtering rules set to be equivalent to those used by the preprocessing SW (``SW rules'' in the figure), the filter achieves a significant reduction in the incoming data rate of \(14.5\%\). This figure also shows the rather small contribution to data reduction attained through the setting of detector-specific filtering rules. Establishing a detector-specific rollover limit,  the ``New RO limit'' case, adds \(0.5\%\) to the rate reduction obtained for the ``SW rules'' case, reaching a \(15\%\) rate decrease. Adding a filtering rule to reject noise events characterized by an abnormally high energy, ``New RO limit + up bound'' case, provides a rather small contribution. However, it is still larger than that of the given by the ``New RO limit'' version,  receiving \(17\%\) less data than in the non-filter case. To understand these results we need to bear in mind the coarse TOT distribution shown in Figure~\ref{fig:e_dist}. Apart from the events identified as valid events undergoing the counter rollover effect present in the region \(-1024 < coarse\_TOT < -950\), the highest concentration of events outside the valid energy range is in the area \(0 < coarse\_TOT < -50\), which concentrates the low-energy events that suffer the effect of the ASIC inter-path length differences. This concentration of non-valid events is eliminated when SW equivalent filtering rules are set. Thus the ``SW rules'' filter implementation is the version with the biggest contribution to the rate reduction. In the other two detector-specific rules implementations, the rate reduction is mainly provided by the filtering of non-valid events belonging to one of the five problematic ASICs identified by the system. The larger the number of problematic ASICs present in the system, the greater the rate reduction for these last two filter versions.

\begin{figure}[htb]
	\centering
	\includegraphics[width=.475\textwidth]{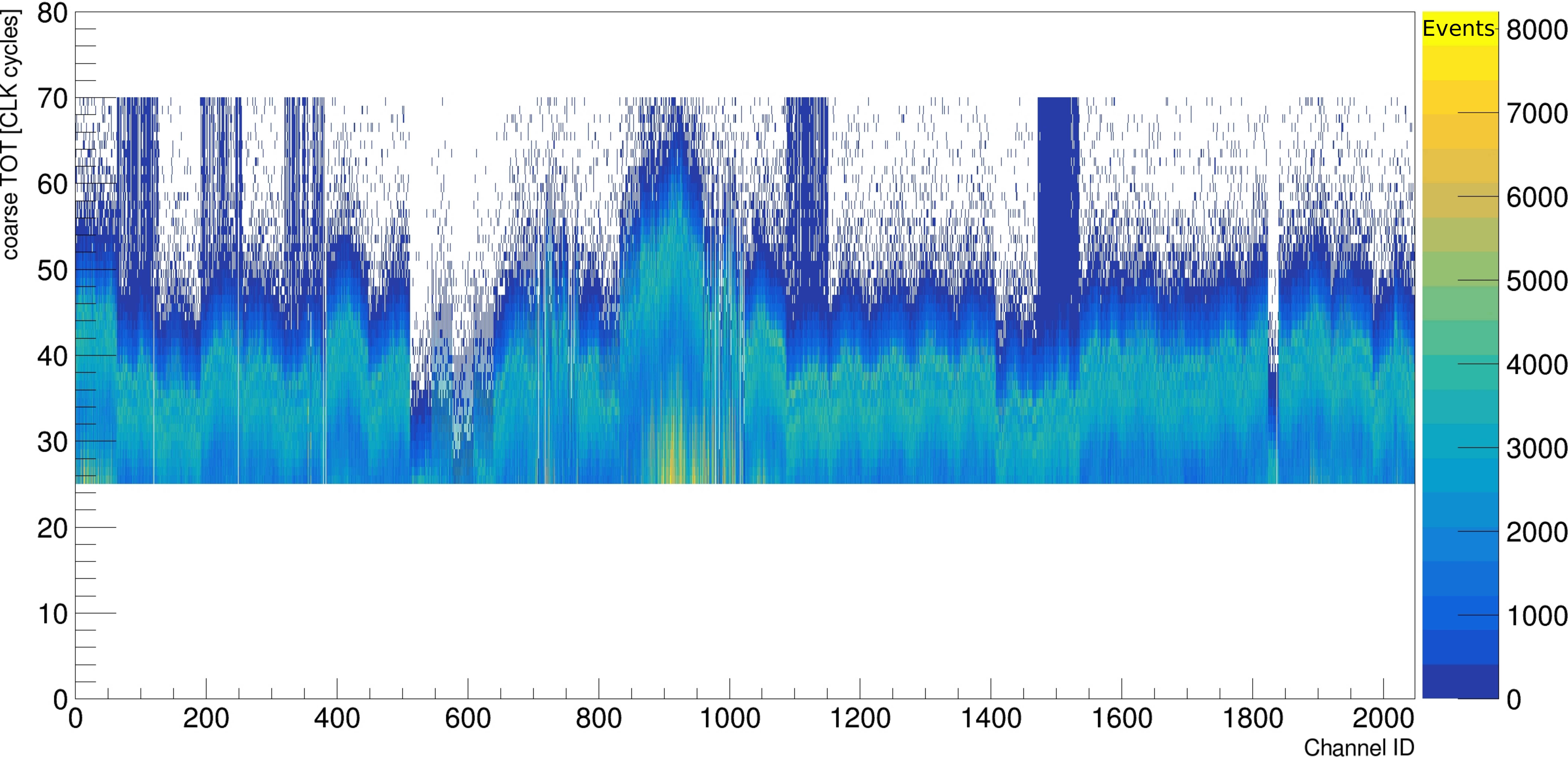}
	\caption{\label{fig:unc_e_ditr} Analysis of TOF-PET scanner energy response. Coarse TOT distribution of the received ASIC events for the 2048 system channels expressed in CLK cycles with 6.25 ns steps. Color scale indicates the number of events. The produced events are the result of a 5-min acquisition placing a Na22 point source centered on the detector ring when the uncalibrated online energy filer is used and setting 25 and 70 CLK cycles as the filtering thresholds on the DAQ front-end modules.}
\end{figure}

\begin{figure}[htb]
	\centering
	\includegraphics[width=.475\textwidth]{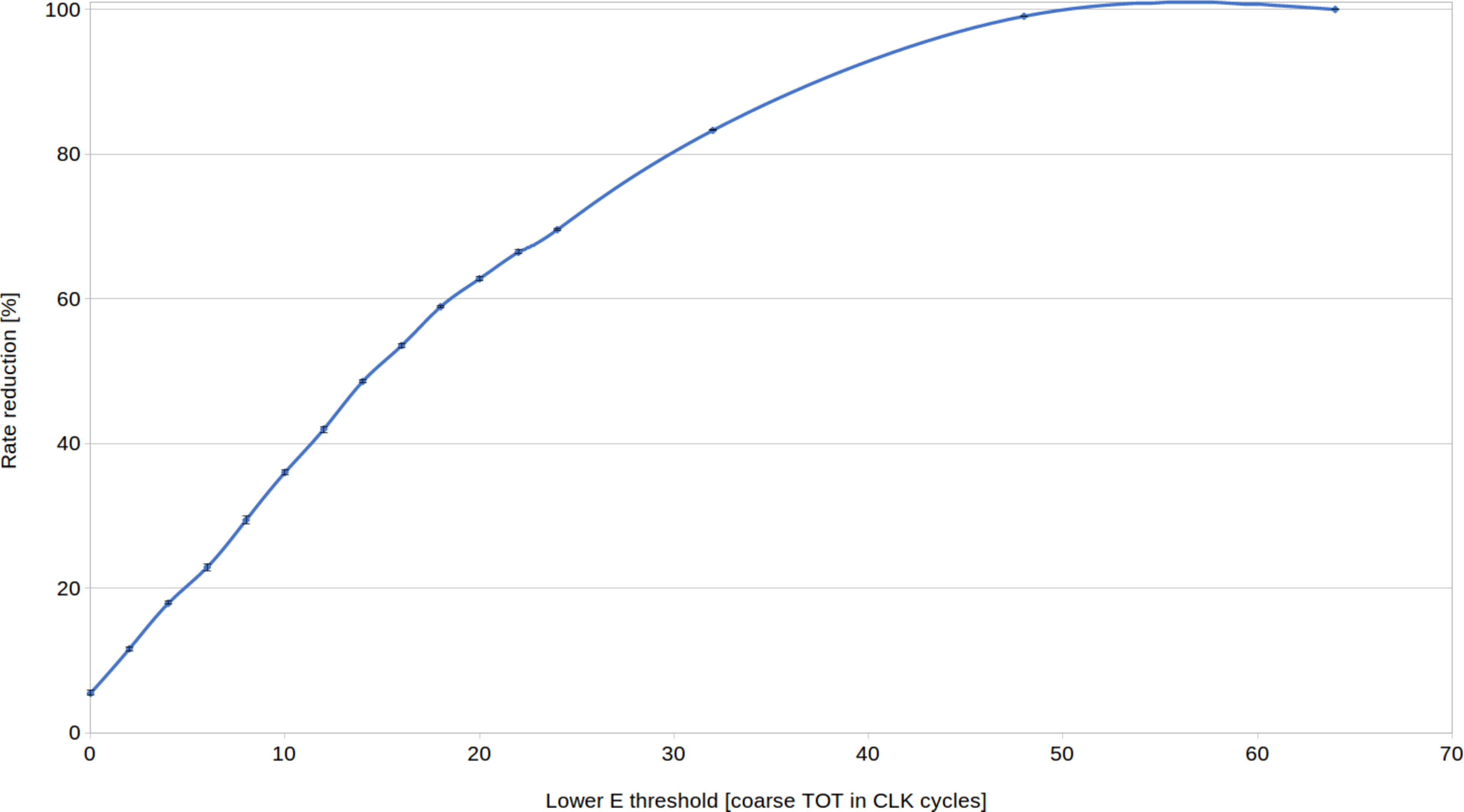}
	\caption{\label{fig:rate_red_uncal_filt} Rate reduction obtained through the increment of the low filter threshold  when applying the uncalibrated online energy filter in the TOF-PET scanner DAQ front-end modules. We express the rate reduction as the percentage relative to the reference non-filtering case.}
\end{figure}

After we determined and set detector-specific hardwired filtering rules, we performed a threshold sweep searching for the best TOT thresholds to apply simultaneously on all channels. The coarse TOT distribution shown in Figure~\ref{fig:unc_e_ditr} illustrates the result of applying the threshold values found to conserve the valid events of every channel, even for those with a lower or higher coarse TOT than the  average. Hence, the resulting lower threshold for the uncalibrated filter was 25 CLK cycles, and the upper threshold was 70 CLK cycles. During the threshold sweep we recorded the system's output data rate, obtaining the rate evolution curve presented in Figure~\ref{fig:rate_red_uncal_filt}. Through this experiment, we observed that by setting the optimum thresholds on the uncalibrated filter, the ones used to get the coarse TOT distribution of Figure~\ref{fig:unc_e_ditr}, we could achieve a \(70\%\) data rate reduction while preserving all the events with a valid energy.

We studied the resources consumption of the filter implementation in the DAQ front-end module FPGA to assess the implementation complexity of this uncalibrated filter version. For the implementation of the filtering module plus the changes in the configuration module, the uncalibrated filter consumes the following FPGA resources for our device under use, Xilinx Kintex7 XC7K160T:
\begin{itemize}
	\item 278 slice LUTs, \(0.27\%\) of the total number of slice LUTs in the device
	\item 510 slice registers, \(0.25\%\) of the total number of slice registers in the device
	\item 119 slices, \(0.47\%\) of the total number of slices in the device
\end{itemize}

\subsection{Calibrated energy filter}\label{sec:res_cal}

\begin{figure}[htb]
	\centering
	\includegraphics[width=.475\textwidth]{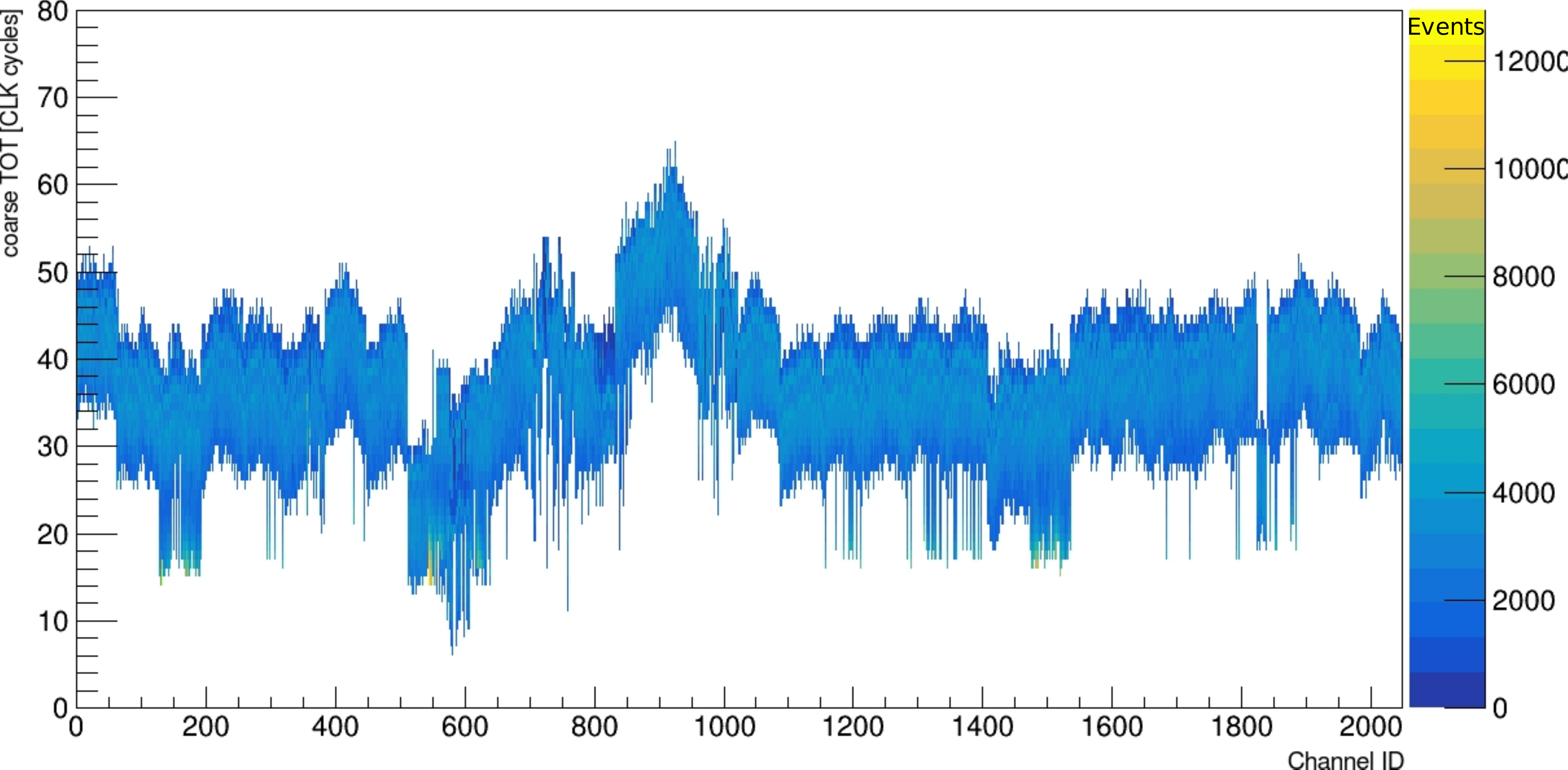}
	\caption{\label{fig:cal_e_ditr} Analysis of TOF-PET scanner energy response. Coarse TOT distribution of the received ASIC events for the 2048 system channels expressed in CLK cycles with 6.25 ns steps. Color scale indicates the number of events. The produced events are the result of a 5-min acquisition placing a Na22 point source centered on the detector ring when the DAQ front-end modules implement the calibrated online energy filter.}
\end{figure}

When we use the calibrated filter version to discriminate non-valid detected events, the resulting coarse TOT distribution of the received events presents variable energy cuts adapted to each channel response as shown in the energy map shown in Figure~\ref{fig:cal_e_ditr}. One of the benefits of using channel-dependent filtering thresholds is the optimum energy-cuts setting that the technique provides. These channel-specific thresholds led to the complete filtering of all events with an energy outside the valid energy range even for malfunctioning ASICs. Observation of the coarse TOT distribution in Figure~\ref{fig:cal_e_ditr} reveals that it is not possible to identify the problematic ASICs present on the system by inspection of their energy spaces, they provide an adequate energy response after filtering. When the filtering process is optimized, for every channel, the DAQ front-end modules only retransmit the gamma events whose energy is within the specified energy window.

\begin{figure}[htb]
	\centering
	\includegraphics[width=.475\textwidth]{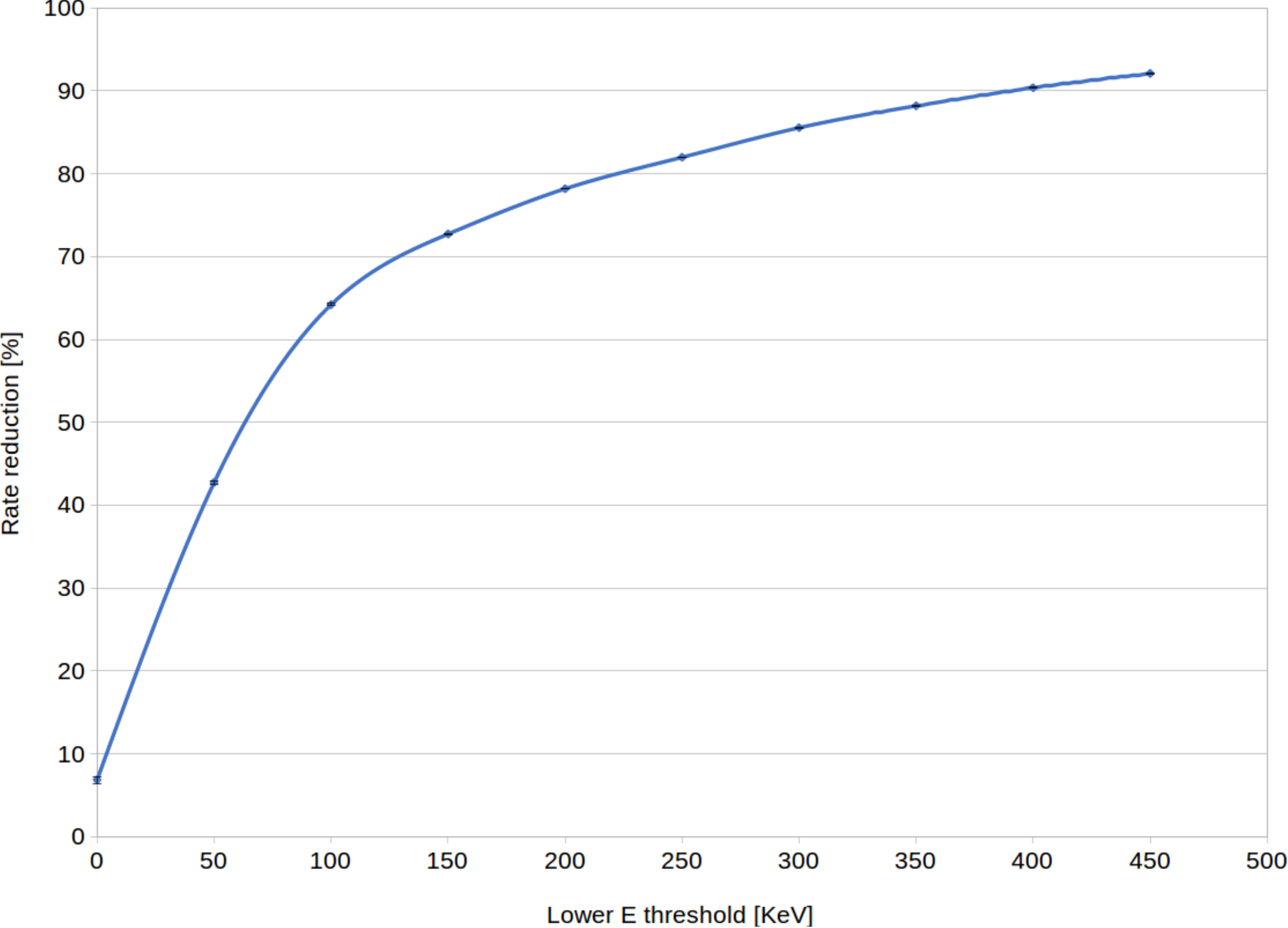}
	\caption{\label{fig:rate_red_cal_filt} Rate reduction obtained through the increment of the low filter threshold  when the DAQ front-end modules implement the calibrated online energy filter on the TOF-PET scanner DAQ front-end modules. We express the rate reduction as percentages relative to the reference non filtering case.}
\end{figure}

Under these conditions we performed a threshold sweep analysis to find the best
energy threshold value for the current system while studying the effect on the data
rate reduction. The evolution curve presented in Figure~\ref{fig:rate_red_cal_filt} shows the progression of the data reduction as we increase the lower threshold. The higher concentration of events with low energies is notable given the reverse exponential tendency of the curve. This result is in agreement with the energy distribution shown in the coarse TOT distribution in
Figure~\ref{fig:e_dist}. The detailed view of the region of interest in the
lower plot in Figure~\ref{fig:e_dist} shows that the lower energy range has the highest concentration of events. Once the low threshold surpasses 100 keV, the evolution on the rate reduction is not so steep. System simulations indicate that the low energy cut value that represents the best compromise between system sensitivity and non-valid event reduction is 225 keV for the EndoTOFPET-US system \cite{Zvolsky:398074}. When we set a lower filter threshold equal to 225 keV, we obtained an \(80\%\) data rate reduction as seen in the evolution curve shown in Figure~\ref{fig:rate_red_cal_filt}.

To analyze the complexity of the calibrated filter version, we studied the resources it consumes when we implement it on the target FPGA. The FPGA utilization report reveals that there is not a significant increment in the resource demand of this version in comparison with the uncalibrated version. For the implementation of the filtering module plus the changes in the configuration module, the calibrated filter consumes the following FPGA resources for the device under use, Xilinx Kintex7 XC7K160T:

\begin{itemize}
	\item 292 slice LUTs, \(0.29\%\) of the total number of slice LUTs in the device
	\item 538 slice registers, \(0.27\%\) of the total number of slice registers in the device
	\item 133 slices, \(0.52\%\) of the total number of slices in the device
	\item 1 block RAM tile, \(0.31\%\) of the total number of block RAM tiles in the device
\end{itemize}

There is a negligible increment of the slice resource needs in comparison to the
uncalibrated filter version. The significant resource consumption difference
between the two filter versions is due to the implementation of the LUT to store
the filtering thresholds of every channel. The implementation of this memory
block consumes one FPGA block RAM tile.

\begin{figure}[htb]
	\centering
	\includegraphics[width=.475\textwidth]{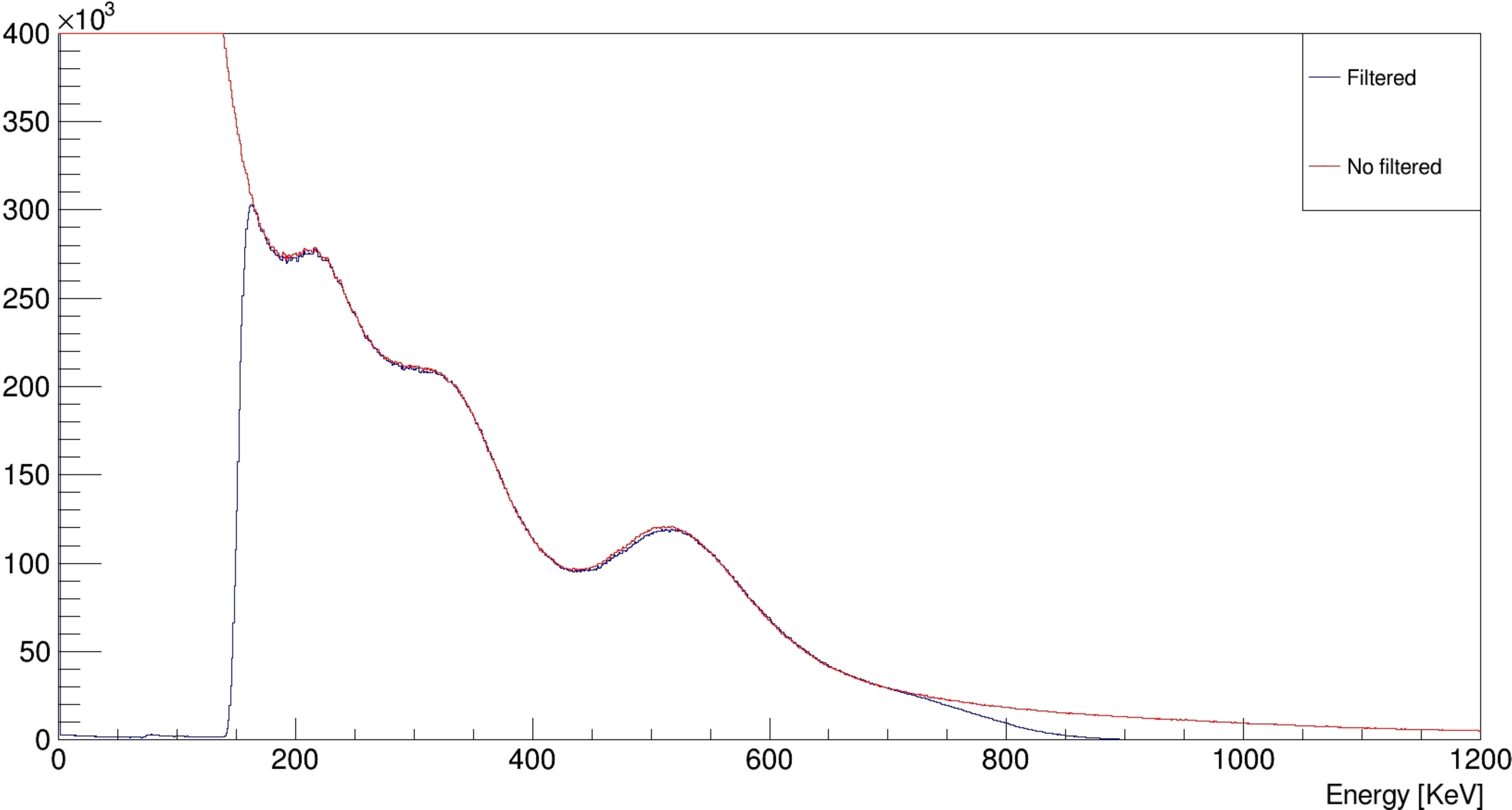}
	\caption{\label{fig:e_sp_comp}  Analysis of TOF-PET scanner energy response. Energy spectra produced by the received ASIC events. The received events are the result of a 5-min acquisition placing a Na22 point source centered on the detector when the DAQ front-end modules implement no filtering at all (red curve) or when the DAQ applies the calibrated online energy filter (blue curve).}
\end{figure}

Despite the indications of proper energy discrimination in the coarse TOT distribution shown in Figure~\ref{fig:rate_red_cal_filt}, we needed to verify that that the usage of the online energy filter does not alter the performance of the system. To ensure that the filtering process is transparent to the system performance, we compared the resulting energy and time resolutions with the original detector resolutions. The energy spectra comparison of Figure~\ref{fig:e_sp_comp} shows the original energy spectrum of the unfiltered data in red and that of the filtered case in blue. For this experiment we set a 150 keV low threshold and a 900 keV upper threshold. The filtered energy spectrum in Figure~\ref{fig:e_sp_comp} demonstrates that proper threshold setting was achieved through the system calibration information used for the experiment; only events with energies in the 150 to 900 keV range passed the filtering stage. The filter threshold setting accuracy is purely determined by the precision of the energy calibration process. For those systems whose calibration process places the optimization focus on a small energy range close to the 511 keV photo-peak, only thresholds set near this value will be accurate. In this case, the further the thresholds are from the photo-peak, the poorer the accuracy will be. Moreover, given the fact that the relationship between coarse TOT and energy is exponential, the resolution is better for lower energies than for higher energies. The energy spectra comparison indicate that when the DAQ front-end modules include the filter, it preserves the original energy resolution because the energy distribution in the photo-peak is the same for both cases. That is, we obtain the same Gaussian fit curve in the 511 keV photo-peak for both energy spectra.

\begin{figure}[htb]
	\centering
	\includegraphics[width=.475\textwidth]{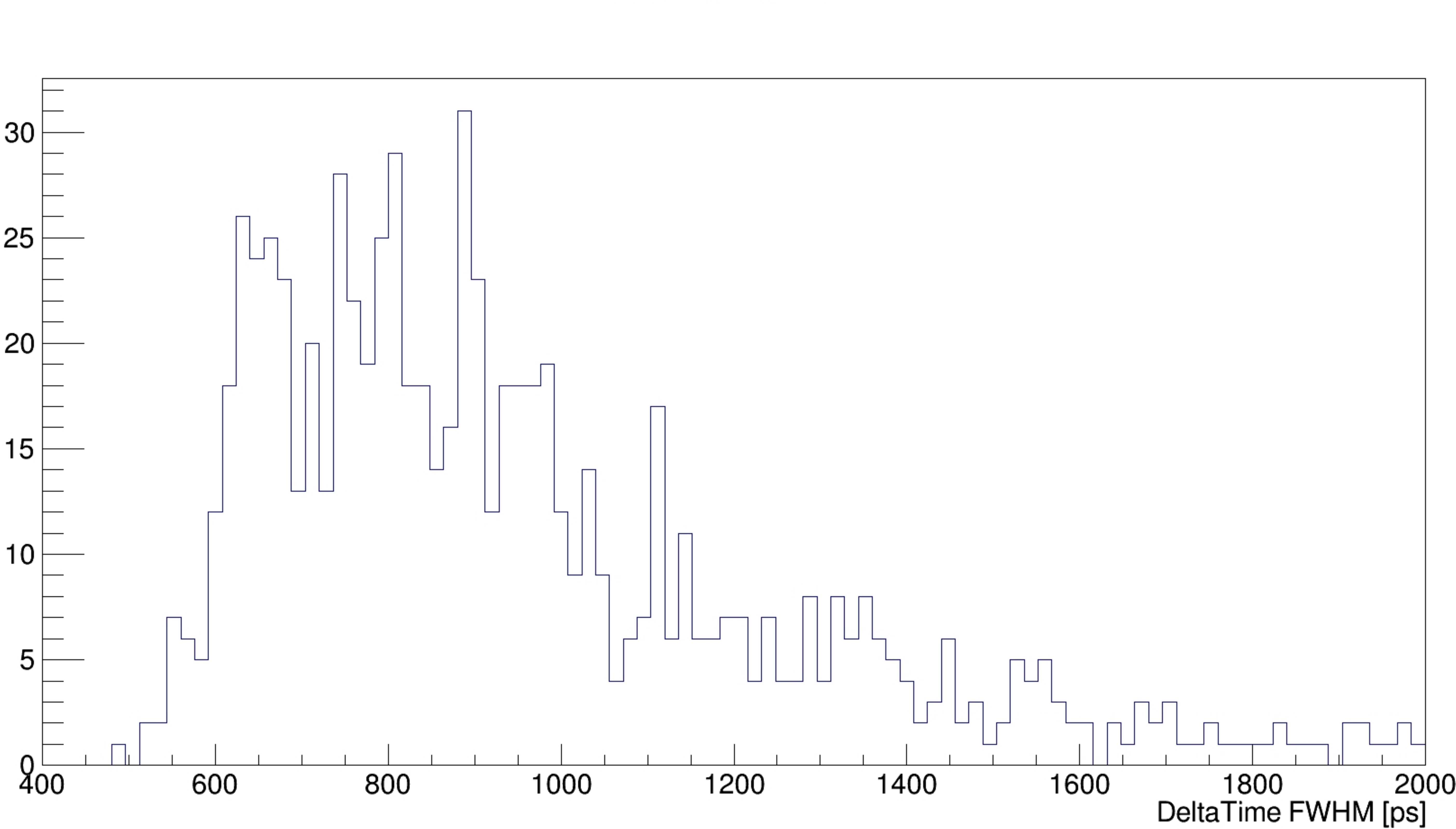}
	\includegraphics[width=.475\textwidth]{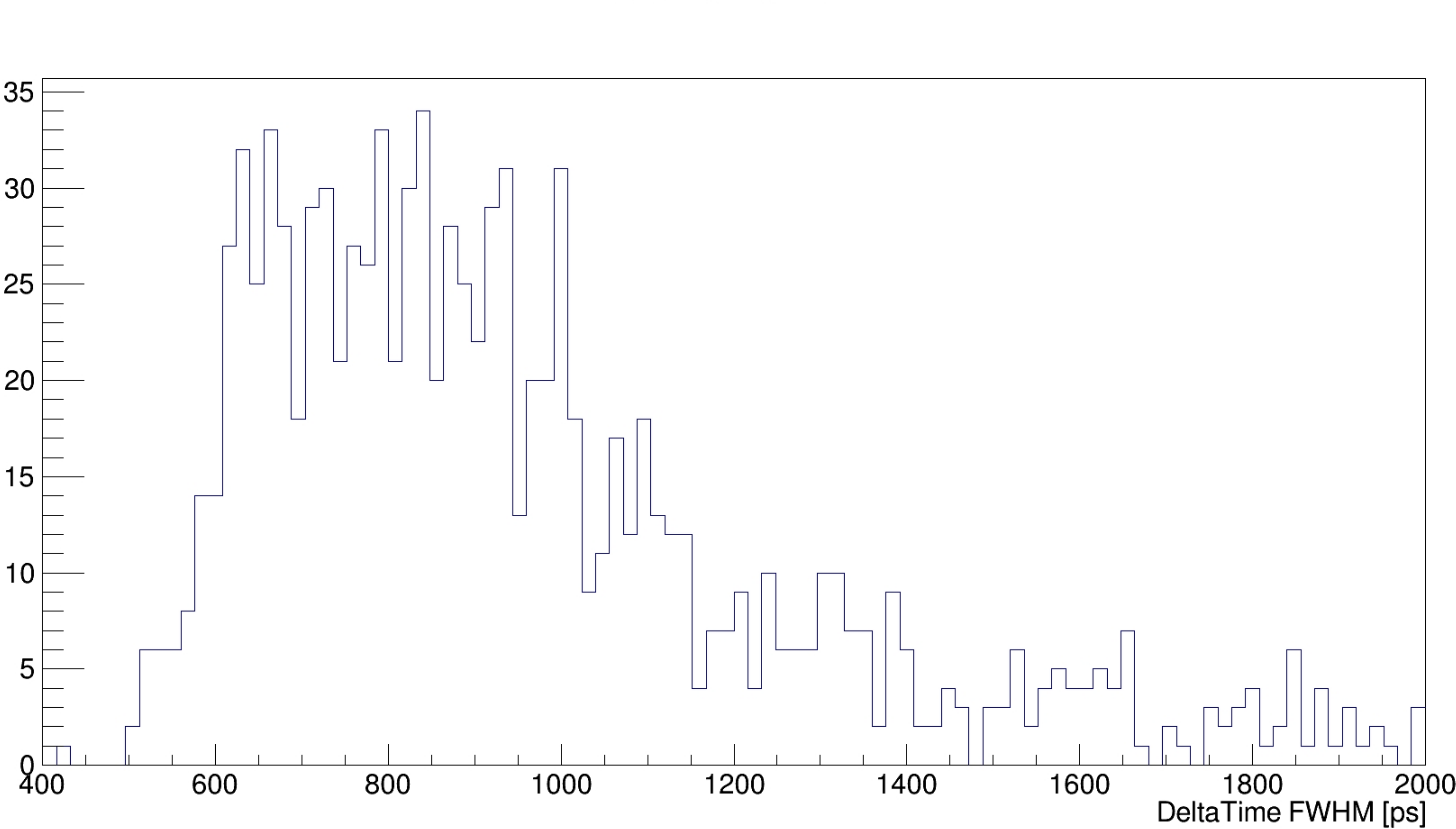}
	\caption{\label{fig:ctr_dist_comp}  Analysis of TOF-PET scanner time response. CTR distributions produced by the received ASIC events. The received events are the result of a 5-min acquisition placing a Na22 point source centered on the detector when the DAQ front-end modules implement no filtering at all (upper graph) and when the DAQ applies the calibrated online energy filter (lower plot).}
\end{figure}

We have demonstrated how the filter does not disturb the system's energy
resolution, but we still need to prove that the system maintains its temporal
properties as well. For this purpose, we perform a comparison of the obtained
coincidence time resolution (CTR) distributions. The CTR distribution shown in the
upper plot in Figure~\ref{fig:ctr_dist_comp} belongs to the non-filtering case and the lower plot displays the CTR distribution obtained when the DAQ front-end modules apply the calibrated energy filter with its lower and upper thresholds set to 150 keV and 900 keV, respectively. The timing results shown in Figure~\ref{fig:ctr_dist_comp} prove the preservation of the timing properties of the system during the filtering process. For both cases, most of the recorded lines-of-response present CTR values around 800 ps full width half maximum (FWHM). This is a poor time resolution for a TOF-PET system and is not in agreement with the previous time resolution reported for this system \cite{1748-0221-11-12-P12003}. The reason for having a different time resolution in this experiment is the need to deactivate the SW post-processing which corrects the event timestamps offline. Only by making use of this post-processing correction, the system may achieve the average CTR values of 375 ps FWHM previously reported for this system.

The ASIC measures the time of arrival of events by reading the charge deposited in a capacitor. Due to a problem in the PETsys Electronics TOFPETv1 ASIC design, this capacitor presents a current leakage; thus, the charge read from the capacitor does not corresponds to the charge associated with its time of arrival. PETsys Electronics characterized this leakage. Thus, knowing the time between consecutive events, the post-processing script can compensate the charge read from the leaking capacitor. Consequently, on systems making use of the PETsys Electronics TOFPETv1 ASIC, the post-processing script needs the time information of the previous event to correct the timestamp of the valid current gamma event independently of the validity of the previous event. As a result, when the DAQ discards a non-valid event, the system loses the
correction information needed for the subsequent event. Given this particularity
of the PETsys Electronics TOFPETv1 ASIC, the online energy filter will only preserve the time properties of the system before applying the SW post-processing time
correction.

\section{Discussion}\label{sec:disc}

Considering the outcomes reported in the results Section~\ref{sec:res}, we
have demonstrated that the proposed implementations of the online energy filter are
effective solutions to reduce the data rate in highly pixelated TOF-PET systems with digital readout while preserving the system performance in terms of sensitivity and resolution (energy and time). The online energy filter is a lightweight solution capable of setting an energy acceptance window on demand to ease the system BW requirements.

The online energy discrimination technique has proven to be especially useful for systems
that include malfunctioning ASICs, which is common in highly pixelated systems comprising many of these readout chips. For the considered case of usage on the
PETsys Electronics demonstrator, we have seen how these problematic ASICs in the system
produce events over the whole energy range. The use of the online energy filter will
contribute to the solution of the BW issue by filtering the events of those ASICs presenting a non-valid energy value.

Our experimental results demonstrate that, for the PETsys Electronics TOFPETv1 ASIC, the biggest contribution to the data rate reduction comes from the filtering of the events in the lowest energy range. The coarse TOT distribution shown in Figure~\ref{fig:e_dist} illustrates the higher concentration of low-energy events, and the coarse TOT histogram shown in Figure~\ref{fig:coarse_tot_dist} confirms that there are five times more events in the lowest energy range than in the range of interest. PETsys Electronics TOFPETv1 ASIC fails to achieve proper adjustment of its triggering scheme; thus, it triggers on the reception of any detected pulse, even those with the lowest energies. Hence, we can conclude that the ASIC is triggered by noise. Not being able to correct this behavior for the current readout chip, the system collects more non-valid events than valid ones. In this system, when the DAQ front-end modules implement an online calibrated filtering stage and a low threshold is set at 225 keV, we achieve an \(80\%\) data reduction. As a consequence, we can conclude that the current system presents a poor signal-to-noise ratio of approximately 1:4; thus, it could greatly benefit from online filtering to remove the ASIC events produced by noise. In general, in systems producing non-valid events along with real gamma photon events, an online energy filter is a good solution for the efficient reduction of the system's output data rate.

Despite the proved efficiency of the filter, we have identified unexpected discrepancies in the system behavior that alter the filter outcome:
\begin{itemize}
	\item PETsys Electronics TOFPETv1 ASIC interdependence between consecutive events. The system must compensate the time information of the current event using the information of the previous event. As a consequence, the DAQ associated with the PETsys Electronics TOFPETv1 ASIC should not apply online energy filtering to preserve the system's time resolution. Although systems readout by PETsys Electronics TOFPETv1 ASIC chips should not apply this technique because of this ASIC particularity, we have been able to prove, on this system, the efficiency of the filtering process by deactivating the SW post-processing compensating the corrupted timestamps. When we disable this process, the	time resolution does not vary when the DAQ implements the filter.
	\item High sensitivity to the energy calibration accuracy and temperature variations. By definition, the calibrated filter implementation is highly dependent on the energy calibration information. This calibration process is not straightforward in systems with thousands of channels. The curve fitting algorithm in search of the optimum TOT-energy curve does not always converge for all the system channels. Furthermore, given the strong exponential TOT-energy relationship, the energy calibration for the higher energy values is not as accurate as that for the	energies in the region of interest. Given these unexpected issues regarding the energy calibration process and the strong dependence of the filter on the calibration information, we found that poor energy calibration will lead to inaccurate positioning of the filter thresholds. Furthermore, SiPMs are highly sensitive to	temperature variation \cite{4774854}, and changes in temperature lead to changes in the TOT-energy relationship. Temperature variations make the present energy calibration information invalid for the new system conditions; thus, the filter will fail to accurately define its filtering limits.
\end{itemize}

Comparing the similar resources consumption and the rate reduction results obtained for the uncalibrated and the calibrated filter implementations, we deduce that, in general, the calibrated filter version achieves better noise reduction on pixelated PET systems than its uncalibrated counterpart. However, given the good rate reduction results obtained using the uncalibrated filter implementation, we conclude that the uncalibrated filter version is a good solution for systems with poor or no energy calibration information.

Although the results obtained by deactivating the post-processing time correction are
conclusive enough to prove the efficiency of the online filtering technique, we
intend to pursue a new test using this technique on a similar system, namely, a highly
pixelated TOF-PET scanner with digital readout. However, this time a different ASIC would be considered. One of the identified future task is the testing of the online energy filter on the second version of the EndoTOFPET-US plate, which makes use of the STiC ASIC\cite{1748-0221-9-02-C02003}. The STiC ASIC does not present any interdependence between consecutive events.

Given the strong dependence of the filter on the energy calibration, which is
commonly inaccurate for values not close to the 511 keV photo-peak, and the variability
of the system response with temperature changes, we envisage a new calibration-independent filter implementation. This version would read all incoming data,
and implement a peak-finder algorithm to identify the position (TOT value) of the
photo-peak for every channel. With this information, we would set the two filtering limits equally spaced on both sides of the photo-peak. This implementation will not let the user specify the filtering energy limits in keV; instead they would be specified as a margin around the photo-peak. This margin setting would enable the control of the systems output data volume while maintaining the system performance if the margin is properly set.

Another upgrade conceived for future work consists of the actual correction of the gamma events timestamps for events undergoing the counter rollover effect. The DAQ front-end modules already process the gamma events time information without modifying it. Correcting these timestamps would be a straightforward process for the DAQ; thus, implementation of such a process would add minimal resource needs to the filter. Making use of such an online procedure would eliminate the necessity of using an offline SW preprocessing mechanism to correct the counter rollover events.

Various PET systems found in literature apply online data filtering techniques that have demonstrated efficiency to reduce the volume of produced data by minimizing the presence of non-useful information \cite{Wu2013, 8069574, 5076010, 7100955}. However, to our knowledge, this is the first time this kind of technique has been applied to a pixelated PET system with a SiPM-ASIC readout scheme. Pixelated PET systems with a high density of digital channels are prone to present noisy channels. The more noisy channels are present in a system, the less BW is available to transmit useful data. Given the ease of access to the digital information of each event on the on-detector DAQ front-end level of PET systems with SiPM-ASIC readout, we have confirmed that energy-based event discrimination is a suitable technique to address the BW limitation issue present in a system with a high count rate, as is the case in highly pixelated PET systems.

\section{Conclusion}\label{sec:conc}

Current trends in PET systems show an increasing interest in highly pixelated systems due to their capability to achieve better spatial resolution than monolithic PET systems. However, pixelated detectors increase the complexity of the DAQ system reading out these detectors. A high density of detecting channels produces a high count rate, which could reach the system BW limit, thus putting the system into saturation and losing valuable information. This work demonstrated that online event discrimination based on energy at the on-detector front-end level is an efficient and lightweight technique to address the BW limitation problem present in highly pixelated PET systems.

Furthermore, through this study we have seen how a PET system readout by the PETsys Electronics TOFPETv1 ASIC would profit from the filtering of the non-valid energy events produced by malfunctioning ASICs. Apart from those non-valid events scattered over the energy space, this ASIC produces the highest concentration of events in the lowest energy range, by filtering those non-useful low-energy events we achieve the largest contribution to the non-valid data rate reduction. In general, any SiPM-ASIC readout scheme generating events with energies out of the range of interest would reduce its BW needs by making use of an online filtering stage that discards those events on the fly. Nonetheless, our analysis also brought to light the timing information corruption issue present in the PETsys Electronics TOFPETv1 ASIC, which creates an unexpected interdependence between consecutive events. Due to this particularity of the chip, we concluded that the DAQ front-end modules should not make use this discriminating technique in systems readout by the PETsys Electronics TOFPETv1 ASIC. Therefore, we plan to conduct further experiments to test the preference of the online energy filter on similar systems readout by other ASICs. Given the results demonstrating the preservation of the system time and energy resolutions, even using the PETsys Electronics TOFPETv1 ASIC when we disable the time correction post-processing mechanism, we conclude that online energy filtering is an effective technique for a pixelated TOF-PET system.

This work provides to the EndoTOFPET-US collaboration a tool to manage the currently overflowing disk usage. We present a data volume control mechanism during singles mode acquisitions where time information is not relevant (e.g. during energy calibration) when the detector plate read by the PETsys Electronics TOFPETv1 ASIC is used. For the plate read by the STiC ASIC, data volume control could maintain system performance in all conditions, even in cases in which time information is needed. From a more global perspective, this study could benefit the PET community in that it has demonstrated the potential and simplicity of applying online processing techniques at the on-detector DAQ level for pixelated TOF-PET systems making use of readout ASICs. Having the digital event information in hand at the DAQ front-end modules, we could apply simple discrimination techniques, such as those analyzed herein, or more complex implementations as the proposed peak-finder discriminator to achieve a calibration-independent solution.

\section{Acknowledgments}\label{sec:ack}
This work, as part of PicoSEC MCNet Project, was supported by a Marie Curie Early Initial Training Network Fellowship of the European Community's Seventh Framework Program [contract number PITN-GA-2011-289355-PicoSEC-MCNet]. Also, EndoTOFPET-US has received funding from the European Union 7th Framework Program,FP7/2007-2013, [Grant Agreement. No. 256984]. This work has also been supported by the Spanish Goverment (project TEC2015-66978-R), Comunidad de Madrid (TOPUS S2013/MIT-3024) and the European Regional Development Funds.

I would like to include a special mention to my ex-colleagues at PETsys Electronics who contributed to the development of some of the tools used in this study: Rui Silva, Miguel Silveira, Carlos Leong and Agostino Di Francesco.

There is a potential conflict of interest given the fact that the following co-authors are employees of PETsys Electronics SA: L. Ferramacho, R. Bugalho, T. Niknejad, J. C. Silva, S. Tavernier, and J. Varela. This relationship with the company commercializing one of the devices under study does not signify any real conflict of interest. We have ensured that the information disclosed in this paper is completely unbiased and objective.

\section*{References}

\bibliographystyle{elsarticle-num}
\bibliography{efilter}

\begin{thebibliography}{10}
\expandafter\ifx\csname url\endcsname\relax
  \def\url#1{\texttt{#1}}\fi
\expandafter\ifx\csname urlprefix\endcsname\relax\def\urlprefix{URL }\fi
\expandafter\ifx\csname href\endcsname\relax
  \def\href#1#2{#2} \def\path#1{#1}\fi

\bibitem{MOSES2011S236}
W.~W. Moses,
  \href{http://www.sciencedirect.com/science/article/pii/S0168900210026276}{{Fundamental
  limits of spatial resolution in PET}}, Nuclear Instruments and Methods in
  Physics Research Section A: Accelerators, Spectrometers, Detectors and
  Associated Equipment 648 (2011) S236 -- S240.
\newblock \href {http://dx.doi.org/https://doi.org/10.1016/j.nima.2010.11.092}
  {\path{doi:https://doi.org/10.1016/j.nima.2010.11.092}}.
\newline\urlprefix\url{http://www.sciencedirect.com/science/article/pii/S0168900210026276}

\bibitem{7843591}
C.~Ritzer, P.~Hallen, D.~Schug, V.~Schulz, {Intercrystal Scatter Rejection for
  Pixelated PET Detectors}, IEEE Transactions on Radiation and Plasma Medical
  Sciences 1~(2) (2017) 191--200.
\newblock \href {http://dx.doi.org/10.1109/TNS.2017.2664921}
  {\path{doi:10.1109/TNS.2017.2664921}}.

\bibitem{Wu2013}
X.~Wu, J.~Zhu, M.~Niu, Z.~Hu, Q.~Xie, P.~Xiao, {Online parameter calibration
  for energy discrimination in trans-PET}, IEEE Nuclear Science Symposium
  Conference Record (2013) 3--5\href
  {http://dx.doi.org/10.1109/NSSMIC.2013.6829181}
  {\path{doi:10.1109/NSSMIC.2013.6829181}}.

\bibitem{8069574}
P.~Conde, A.~Iborra, A.~J. Gonz{\'{a}}lez, A.~Aguilar, E.~D{\'{i}}az-Caballero,
  J.~J. Garc{\'{i}}a-Garrigos, A.~Gonz{\'{a}}lez-Montoro, D.~Grau-Ru{\'{i}}z,
  S.~S{\'{a}}nchez, L.~Hern{\'{a}}ndez, P.~Bellido, L.~Moliner, J.~P. Rigla,
  M.~J. Rodr{\'{i}}guez-{\'{A}}lvarez, F.~S{\'{a}}nchez, M.~Seimetz,
  A.~Soriano, L.~F. Vidal, J.~M. Benlloch, {Noise rejection in monolithic PET
  detectors}, in: 2016 IEEE Nuclear Science Symposium, Medical Imaging
  Conference and Room-Temperature Semiconductor Detector Workshop
  (NSS/MIC/RTSD), 2016, pp. 1--5.
\newblock \href {http://dx.doi.org/10.1109/NSSMIC.2016.8069574}
  {\path{doi:10.1109/NSSMIC.2016.8069574}}.

\bibitem{5076010}
J.~D. Leroux, M.~A. Tetrault, D.~Rouleau, C.~M. Pepin, J.~B. Michaud,
  J.~Cadorette, R.~Fontaine, R.~Lecomte, {Time Discrimination Techniques Using
  Artificial Neural Networks for Positron Emission Tomography}, IEEE
  Transactions on Nuclear Science 56~(3) (2009) 588--595.
\newblock \href {http://dx.doi.org/10.1109/TNS.2009.2021428}
  {\path{doi:10.1109/TNS.2009.2021428}}.

\bibitem{7100955}
M.~A. T{\'{e}}trault, A.~C. Therrien, {\'{E}}.~D. Lamy, A.~Boisvert,
  R.~Fontaine, J.~F. Pratte, {Dark Count Impact for First Photon Discriminators
  for SPAD Digital Arrays in PET}, IEEE Transactions on Nuclear Science 62~(3)
  (2015) 719--726.
\newblock \href {http://dx.doi.org/10.1109/TNS.2015.2420795}
  {\path{doi:10.1109/TNS.2015.2420795}}.

\bibitem{Aubry2013}
N.~Aubry, E.~Auffray, F.~B. Mimoun, N.~Brillouet, R.~Bugalho, E.~Charbon,
  O.~Charles, D.~Cortinovis, P.~Courday, a.~Cserkaszky, C.~Damon, K.~Doroud,
  J.~M. Fischer, G.~Fornaro, J.~M. Fourmigue, B.~Frisch, B.~F{\"{u}}rst,
  J.~Gardiazabal, K.~Gadow, E.~Garutti, C.~Gaston, a.~Gil-Ortiz, E.~Guedj,
  T.~Harion, P.~Jarron, J.~Kabadanian, T.~Lasser, R.~Laugier, P.~Lecoq,
  D.~Lombardo, S.~Mandai, E.~Mas, T.~Meyer, O.~Mundler, N.~Navab,
  C.~Ortig{\~{a}}o, M.~Paganoni, D.~Perrodin, M.~Pizzichemi, J.~O. Prior,
  T.~Reichl, M.~Reinecke, M.~Rolo, H.~C. Schultz-Coulon, M.~Schwaiger, W.~Shen,
  a.~Silenzi, J.~C. Silva, R.~Silva, I.~S. Schweiger, R.~Stamen, J.~Traub,
  J.~Varela, V.~Veckalns, V.~Vidal, J.~Vishwas, T.~Wendler, C.~Xu, S.~Ziegler,
  M.~Zvolsky,
  \href{http://stacks.iop.org/1748-0221/8/i=04/a=C04002}{{EndoTOFPET-US: a
  novel multimodal tool for endoscopy and positron emission tomography}},
  Journal of Instrumentation 8~(04) (2013) C04002.
\newblock \href {http://dx.doi.org/10.1088/1748-0221/8/04/C04002}
  {\path{doi:10.1088/1748-0221/8/04/C04002}}.
\newline\urlprefix\url{http://stacks.iop.org/1748-0221/8/i=04/a=C04002}

\bibitem{10.1007/978-3-319-00846-2_112}
C.~Zorraquino, {EndoTOFPET-US a High Resolution Endoscopic PET-US Scanner Used
  for Pancreatic and Prostatic Clinical Exams}, in: L.~M. {Roa Romero} (Ed.),
  XIII Mediterranean Conference on Medical and Biological Engineering and
  Computing 2013, Springer International Publishing, Cham, 2014, pp. 451--454.

\bibitem{Frisch2013}
B.~Frisch, \href{http://dx.doi.org/10.1016/j.nima.2013.05.027}{{Combining
  endoscopic ultrasound with Time-Of-Flight PET: The EndoTOFPET-US Project}},
  Nuclear Instruments and Methods in Physics Research, Section A: Accelerators,
  Spectrometers, Detectors and Associated Equipment 732 (2013) 577--580.
\newblock \href {http://dx.doi.org/10.1016/j.nima.2013.05.027}
  {\path{doi:10.1016/j.nima.2013.05.027}}.
\newline\urlprefix\url{http://dx.doi.org/10.1016/j.nima.2013.05.027}

\bibitem{Bugalho2013}
R.~Bugalho, C.~Gaston, M.~D. Rolo, J.~C. Silva, R.~Silva, J.~Varela,
  \href{http://stacks.iop.org/1748-0221/8/i=02/a=C02049?key=crossref.53fa6832c5d9e999944e3bb8145d2cdb}{{EndoTOFPET-US
  data acquisition system}}, Journal of Instrumentation 8~(02) (2013)
  C02049--C02049.
\newblock \href {http://dx.doi.org/10.1088/1748-0221/8/02/C02049}
  {\path{doi:10.1088/1748-0221/8/02/C02049}}.
\newline\urlprefix\url{http://stacks.iop.org/1748-0221/8/i=02/a=C02049?key=crossref.53fa6832c5d9e999944e3bb8145d2cdb}

\bibitem{7407512}
C.~Zorraquino, R.~Bugalho, M.~Rolo, J.~C. Silva, V.~Vecklans, R.~Silva,
  C.~Ortig{\~{a}}o, J.~A. Neves, S.~Tavernier, P.~Guerra, A.~Santos, J.~Varela,
  {Asymmetric Data Acquisition System for an Endoscopic PET-US Detector}, IEEE
  Transactions on Nuclear Science 63~(1) (2016) 213--221.
\newblock \href {http://dx.doi.org/10.1109/TNS.2016.2514600}
  {\path{doi:10.1109/TNS.2016.2514600}}.

\bibitem{1748-0221-8-02-C02050}
M.~D. Rolo, R.~Bugalho, F.~Gon{\c{c}}alves, G.~Mazza, A.~Rivetti, J.~C. Silva,
  R.~Silva, J.~Varela,
  \href{http://stacks.iop.org/1748-0221/8/i=02/a=C02050}{{TOFPET ASIC for PET
  applications}}, Journal of Instrumentation 8~(02) (2013) C02050.
\newline\urlprefix\url{http://stacks.iop.org/1748-0221/8/i=02/a=C02050}

\bibitem{1748-0221-11-12-P12003}
T.~Niknejad, S.~Setayeshi, S.~Tavernier, R.~Bugalho, L.~Ferramacho, A.~D.
  Francesco, C.~Leong, M.~D. Rolo, M.~Shamshirsaz, J.~C. Silva, R.~Silva,
  M.~Silveira, C.~Zorraquino, J.~Varela,
  \href{http://stacks.iop.org/1748-0221/11/i=12/a=P12003}{{Validation of a
  highly integrated SiPM readout system with a TOF-PET demonstrator}}, Journal
  of Instrumentation 11~(12) (2016) P12003.
\newline\urlprefix\url{http://stacks.iop.org/1748-0221/11/i=12/a=P12003}

\bibitem{ORITA2011S24}
T.~Orita, H.~Takahashi, K.~Shimazoe, T.~Fujiwara, S.~Boxuan,
  \href{http://www.sciencedirect.com/science/article/pii/S0168900211000738}{{A
  new pulse width signal processing with delay-line and non-linear circuit (for
  ToT)}}, Nuclear Instruments and Methods in Physics Research Section A:
  Accelerators, Spectrometers, Detectors and Associated Equipment 648 (2011)
  S24 -- S27.
\newblock \href {http://dx.doi.org/https://doi.org/10.1016/j.nima.2011.01.023}
  {\path{doi:https://doi.org/10.1016/j.nima.2011.01.023}}.
\newline\urlprefix\url{http://www.sciencedirect.com/science/article/pii/S0168900211000738}

\bibitem{Zvolsky:398074}
M.~Zvolsky, \href{https://bib-pubdb1.desy.de/record/398074}{{S}imulation,
  {I}mage {R}econstruction and {S}i{PM} {C}haracterisation for a {N}ovel
  {E}ndoscopic {P}ositron {E}mission {T}omography {D}etector}, Dissertation,
  Universität Hamburg, Hamburg, dissertation, Universität Hamburg, 2017
  (2017).
\newblock \href {http://dx.doi.org/10.3204/PUBDB-2017-13685}
  {\path{doi:10.3204/PUBDB-2017-13685}}.
\newline\urlprefix\url{https://bib-pubdb1.desy.de/record/398074}

\bibitem{4774854}
M.~Ramilli, Characterization of sipm: Temperature dependencies, in: 2008 IEEE
  Nuclear Science Symposium Conference Record, 2008, pp. 2467--2470.
\newblock \href {http://dx.doi.org/10.1109/NSSMIC.2008.4774854}
  {\path{doi:10.1109/NSSMIC.2008.4774854}}.

\bibitem{1748-0221-9-02-C02003}
T.~Harion, K.~Briggl, H.~Chen, P.~Fischer, A.~Gil, V.~Kiworra, M.~Ritzert,
  H.~C. Schultz-Coulon, W.~Shen, V.~Stankova,
  \href{http://stacks.iop.org/1748-0221/9/i=02/a=C02003}{{STiC — a mixed mode
  silicon photomultiplier readout ASIC for time-of-flight applications}},
  Journal of Instrumentation 9~(02) (2014) C02003.
\newline\urlprefix\url{http://stacks.iop.org/1748-0221/9/i=02/a=C02003}

\end{thebibliography}

\end{document}